\documentclass{elsarticle}

\usepackage{lineno,hyperref}
\usepackage{amsmath,amssymb,amsfonts}
\usepackage{algorithmic}
\usepackage{graphicx}
\usepackage{textcomp}
\usepackage[dvipsnames]{xcolor}

\usepackage{pifont}

\usepackage{xspace}
\usepackage{booktabs} \usepackage[utf8]{inputenc}
\usepackage{amsmath,amssymb,stmaryrd}
\usepackage{paralist}
\usepackage{flushend}
\usepackage[justification=centering]{subfig}
\usepackage{soul}
\usepackage{balance}
\usepackage{makecell}
\usepackage{framed}
\colorlet{shadecolor}{blue!10}
\usepackage{verbatim}
\usepackage{multirow}
\usepackage{adjustbox}
\usepackage{enumitem}

\usepackage{listings}
\newcommand{\ie}{i.e.,\xspace}
\newcommand{\eg}{e.g.,\xspace}

\newcommand{\etal}{\emph{et~al.}\xspace}
\newcommand{\secref}[1]{Section~\ref{#1}\xspace}
\newcommand{\figref}[1]{Figure~\ref{#1}\xspace}

\newcommand{\tabref}[1]{Table~\ref{#1}\xspace}

\usepackage{mdframed}
\usepackage{ntheorem}
\theoremstyle{nonumberplain}
\newmdtheoremenv[linecolor=blue,
  linewidth=0pt,rightline=false,
  leftline=false]{figrev}{}

\usepackage[numbers]{natbib}
\usepackage{url}
\usepackage{xargs}  
\modulolinenumbers[5]

\makeatletter
\def\ps@pprintTitle{%
  \let\@oddhead\@empty
  \let\@evenhead\@empty
  \let\@oddfoot\@empty
  \let\@evenfoot\@oddfoot
}
\makeatother

\usepackage{mwe}
\newlength{\tempheight}

\newcommand{\rowname}[1]{\rotatebox{90}{\makebox[\tempheight][t]{\scriptsize{#1}}}}

\usepackage{todonotes}
\definecolor{OliveGreen}{rgb}{0.25,0.5,0.35} \definecolor{Plum}{rgb}{0.25,0.35,0.75} \definecolor{Red}{rgb}{1,0.25,0}

\newboolean{showcomments}
\setboolean{showcomments}{true}

\ifthenelse{\boolean{showcomments}}
  {\newcommand{\nb}[2]{
    \fbox{\bfseries\sffamily\scriptsize#1}
    {\sf\small$\blacktriangleright$\textit{#2}$\blacktriangleleft$}
   }
  }
  {\newcommand{\nb}[2]{}
  }
\newcommand\DDP[1]{\textcolor{Fuchsia}{\nb{DANIELE}{#1}}}
\newcommand\RE[1]{\textcolor{purple}{\nb{ROMI}{#1}}}

\newcommand\VIC[1]{\textcolor{BrickRed}{\nb{VITTORIO}{#1}}}

\begin{document}

\begin{frontmatter}

\title{A Model-driven Approach for Continuous Performance Engineering in Microservice-based Systems\tnoteref{t1}}
\tnotetext[t1]{This research was supported by the AIDOaRt project (ECSEL-JU program - grant agreement n.101007350).}

\author{Vittorio Cortellessa}
\ead[email]{vittorio.cortellessa@univaq.it}
\author{Daniele {Di Pompeo}}
\ead[email]{daniele.dipompeo@univaq.it}
\author{Romina Eramo}
\ead[email]{romina.eramo@univaq.it}
\author{Michele Tucci\corref{mycorrespondingauthor}}
\ead[email]{michele.tucci@univaq.it}

\cortext[mycorrespondingauthor]{Corresponding author}
\address{DISIM, University of L'Aquila, \\ Via Vetoio, L'Aquila, ITALY}

\begin{abstract}

\noindent
Microservices are quite widely impacting on the software industry in recent years.  Rapid evolution and continuous deployment represent specific benefits of microservice-based systems, but they may have a significant impact on non-functional properties like performance. Despite the obvious relevance of this property, there is still a lack of systematic approaches that explicitly take into account performance issues in the lifecycle of microservice-based systems. 

In such a context of evolution and re-deployment, Model-Driven Engineering techniques can provide major support to various software engineering activities, and in particular they can allow managing the relationships between a running system and its architectural model. 

In this paper, we propose a model-driven integrated approach that exploits traceability relationships between the monitored data of a microservice-based running system and its architectural model to derive recommended refactoring actions that lead to performance improvement.
The approach has been applied and validated on two microservice-based systems, in the domain of e-commerce and ticket reservation, respectively, whose architectural models have been designed in UML profiled with MARTE.

\end{abstract}


\end{frontmatter}


\section{Introduction}\label{sec:intro}

\noindent
Microservices have become a popular style for architecting a software system as a suite of small services, and they are nowadays adopted by many key technological players such as Netflix, Amazon, and Google.  
Major benefits of a microservice-based architecture are that it ensures loose coupling, and it supports rapid evolution and continuous deployment. In addition, having a large set of independently developed services helps in terms of developer productivity, scalability, and maintainability. 
Contextually, the rapidly growing complexity of software systems has forced practitioners to use and investigate different development techniques to tackle advances in productivity and quality. To this extent, software engineering needs to relay on automated approaches to keep low the development costs while tackling the rapid changes of software capabilities that may considerably impact non-functional properties like performance.

In order to manage software complexity, ever more companies in the last two decades have embedded Model-Driven Engineering (MDE)~\cite{MDE} approaches in their processes, with the perceived benefit of enabling developers to work at a higher level of abstraction and to rely on automation throughout the development process.  Nevertheless, MDE solutions need to be further developed to scale up for real-life industrial projects~\cite{MDEDeRun18}. To this intent, one of the major challenges is to work on achieving a more efficient integration between the design and runtime aspects of systems.
For instance, through observation and instrumentation, logs and metrics can be collected and related to the original software design in order to comprehend, extrapolate and analyze the inner behavior of a running software system~\cite{CitoLBKMG18}.

In this context, non-functional properties (e.g., performance, power consumption or memory footprint) are becoming ever more relevant for the success of a software application, and the early identification of problems induces lower cost solutions~\cite{WoodsideFP07}. 
On one side, in model-based software performance engineering, a number of approaches have been proposed for detecting and removing performance problems in software models. Some techniques are based on the concept of performance antipattern, which characterizes bad design practices that may jeopardize software performance, along with possible refactoring actions aimed to remove them~\cite{Cortellessa2013}.
On the other side, methods and tools have been proposed for monitoring system execution and measuring performance of running systems. However, many of them do not envisage a solid integration with architectural design models~\cite{MDEDeRun18}. 
Instead, one of the main benefits in adopting model-based performance evaluation is the ability to conduct analysis (\eg what-if analysis) that would be expensive on a real system, such as to analyze the system behavior when exposed to different workloads, or to analyze the performance sensitivity to system parameter variations. 
Basing on a solid connection between runtime information and architectural design, developers can suggest architectural changes aimed, for example, at meeting performance requirements before the system actually experiments certain scenarios (\eg some specific workloads).

In this paper, we propose a model-driven approach to realize a continuous software engineering loop in microservice-based systems. The approach exploits design-runtime interactions to support designers of microservices in performance analysis and system refactoring tasks. In particular, microservices are monitored by means of distributed tracing, \ie logs are stored in a central location and metrics for all instances of a given service are aggregated to understand the overall state. The observed system behavior at runtime is related to the architectural design to investigate potential performance issues and to design and implement effective system refactoring actions. 

In order to realize our approach, we exploit the specific characteristics of these systems. In fact, each microservice is a separate and autonomous entity that can change independently of each other. This allows us to make a change to a single service and deploy it independently of the rest of the system. In contrast, in monolithic applications any refactoring would impact a large amount of the system, requiring additional coordination among components when making changes; also, in order to release changes, the whole monolithic application should be deployed, implying to manage a higher delta between releases and a higher risk of malfunctioning \cite{Newman15}.

In a previous paper~\cite{ACPET19}, we presented a preliminary version of this approach. It has been realized within Eclipse EMF\footnote{Eclipse Modeling Framework: https://www.eclipse.org/modeling/emf/}. 
It integrates a model-driven framework for
the definition of a traceability model between logs extracted from a running system and an architectural model, which has been realized by means of JTL~\cite{CDEP10}. Basing on feeding an architectural model with runtime monitored performance indices,
we have adopted an end-to-end solution for performance improvement~\cite{ARCELLI2018366}.
This paper extends our previous work as follows: 
\begin{itemize}
    \item[-] We extended the work with the translation of refactoring actions suggested by performance model analysis into refactoring actions applied to the running system, thus closing a continuous performance engineering loop  that was missing in our original work. 
    \item[-] 
We have enabled the detection of performance antipatterns on the basis of performance indices gathered from a running system instead of employing model-based estimated performance indices. In addition, we introduce here the extraction of an additional performance index, i.e. CPU utilization.
    This inevitably leads to more realistic evaluations of performance problems. 
    \item[-] In our previous work, we applied the approach to a single case study, E-Shopper, at the modeling level. In this paper, beside completing the application of the whole performance engineering loop to E-Shopper (i.e., at modeling and running system levels), we also apply the approach on an additional case study, Train Ticket, which is larger than the previous one and it comes from the literature.
\item[-] In this work we present an evaluation of the approach; in particular, we answer to two research questions, which we did not introduce in the former study, that are: (RQ1) Do the proposed model refactoring actions improve the performance of the running system? (RQ2) To what extent does performance antipattern (PA) removal improve the whole performance?
\end{itemize}

As in our previous work, the approach is here applied on microservice-based systems modeled by means of UML~\cite{UML2} profiled with MARTE~\cite{MARTE}, which is the official OMG\footnote{Object Management Group: \url{https://www.omg.org/}} profile for augmenting the UML with quantitative knowledge. In particular, for behavioral aspects we consider Sequence Diagrams, whereas for static aspects we consider Use Case, Component, and Deployment Diagrams.

The rest of the paper is organized as follows: Section~\ref{sec:background} introduces background information about the techniques used in the work; Section~\ref{sec:approach} describes the proposed model-driven approach for continuous performance engineering; in Section~\ref{sec:evaluation} the approach is applied and validated on two realistic microservice-based systems; threats to validity are discussed in Section~\ref{sec:t2v}; related work is presented in Section~\ref{sec:related}, and finally Section~\ref{sec:conclusion} concludes the paper.

 \section{Background}\label{sec:background}

\noindent
In the following, we describe the background of this work in terms of existing techniques that have been adopted.

\subsection{Monitoring infrastructure}\label{subsec:monitoring_techniques}

\noindent
One of the defining characteristics of microservice-based systems is that each service must be independently deployable~\cite{Chen18}.
This aspect favors the independent development of services, but it also makes traditional application monitoring insufficient.
While suitable for monolithic applications, the gathering of logs and metrics of each service does not provide a complete understanding of the system behavior.
This is the main reason for the adoption of distributed tracing as a mean to correlate events generated in individual services with a transaction traversing the entire system.

In this work, we focused on microservices applications deployed on \emph{Docker}\footnote{Docker: \url{https://www.docker.com/}} and developed with \emph{Spring Boot}\footnote{Spring Boot: \url{https://spring.io/projects/spring-boot}} and \emph{Spring Cloud}\footnote{Spring Cloud: \url{https://spring.io/projects/spring-cloud}}. As a consequence, we chose \emph{Spring Cloud Sleuth}\footnote{Spring Cloud Sleuth: \url{https://spring.io/projects/spring-cloud-sleuth}} to implement a distributed tracing solution. In \emph{Spring Cloud Sleuth}, a trace consists of a series of casually related events that are triggered by a request as it moves through a distributed system. These events are called spans and they represent a timed operation occurring in a component. Spans contain references to other spans, which allow a trace to be assembled as a complete workflow. A span contains a set of basic information: the name of the operation, the name of the component providing the operation, the start timestamp and duration (or, alternatively, the finish timestamp), the role of the span in the request and a set of user-defined annotations called tags. Beside basic information, spans generated by \emph{Spring Cloud Sleuth} also contain the IP address and port number of the service, the Java class and method implementing the operation, as well as the unique identifier of the \emph{Spring Cloud} instance.

Once the application is instrumented to produce traces, an infrastructure is necessary to collect and store them.
In our approach, the traces produced by each service during the execution are gathered by the \emph{Zipkin}\footnote{Zipkin: \url{https://zipkin.io/}} distributed tracing system. In turn, \emph{Zipkin} is configured to forward the monitoring data to the distributed database and search engine \emph{Elasticsearch}\footnote{Elasticsearch: \url{https://www.elastic.co/products/elasticsearch}}.

The resources utilization of individual microservices is another important aspect to consider when monitoring system performance. For this work, we selected \emph{perf}\footnote{perf: \url{https://perf.wiki.kernel.org/}} among the wide range of performance analyzing tools. \emph{Perf} is one of the most commonly used performance counter profiling tools on Linux and supports hardware and software performance counters, tracepoints and dynamic probes. We used \emph{perf} to gather accurate CPU utilization for each microservice. We stored such measures in \emph{Elasticsearch}, along with the traces collected by \emph{Zipkin}.

\noindent

\subsection{MDE techniques}\label{subsec:mde}

\noindent
Model Driven Engineering (MDE)~\cite{MDE} leverages intellectual property and business logic from source code into high-level specifications enabling more accurate analyses. In general, an application domain is consistently analyzed and engineered by means of a \emph{metamodel}, \ie a coherent set of interrelated concepts. A model is said to \emph{conform} to a metamodel, meaning that the former is expressed by the concepts encoded in the latter. Constraints are defined at the meta-level, and the consistency relationships between models are guaranteed by means of (bidirectional) model transformations specified on source and target metamodels. With the introduction of model-driven techniques in the software lifecycle, also the analysis of non-functional properties has become effective by means of dedicated tools for the automated assessment of quality attributes~\cite{Cortellessa2011}.

In this work, we used two model-driven frameworks to define model-driven traceability links to relate software architecture and runtime information and to perform performance analysis and model refactoring, respectively. Such frameworks are introduced in the follows.

\subsubsection{JTL}\label{subsec:jtl}
\noindent
JTL (Janus Transformation Language)~\cite{CDEP10} is an Eclipse EMF-based tool realized to maintain consistency between software artifacts\footnote{JTL: \url{https://github.com/MDEGroup/jtl-eclipse}}. Its constraint-based and relational model transformation engine is specifically tailored to support bidirectionality, change propagation and traceability. Within the framework, designers can specify model transformations as bidirectional relationships between elements of two domains (\ie metamodels). The bidirectional engine provides the possibility to apply the transformation rules in both ways, from right to left domains and vice versa.
The JTL transformation mechanism provides a relational semantics relying on Answer Set Programming (ASP)~\cite{GL88} and uses the DLV constraint solver~\cite{DLV} to find consistent solutions.

In this work, we used the JTL traceability mechanism to store relevant details about the linkage between right and left model elements at execution-time~\cite{EPT18}. In MDE, such linkages are often based on the concept of traceability relationships, which may help designers to understand associations and dependencies that exist among heterogeneous models~\cite{PaigeDKFPOZ11,Winkler2010}. 
In particular, a \emph{traceability link} is a relationship between one or more source model elements and one or more target model elements, whereas a \emph{trace model} is a structured set of traceability links, \eg between source and target models. 
Within JTL, traceability links are extrapolated during the transformation execution and made explicit by the framework. In fact, traceability models are maintained as models conforms to the dedicated traceability metamodel, as defined in its Ecore format within EMF. Traceability models can be stored, viewed and manipulated (if needed) by the designer. 

\subsubsection{PADRE}\label{subsec:padre}
\noindent
PADRE (Performance Antipatterns Detection and model REfactoring)~\cite{ARCELLI2018366} is a unified framework that tries to improve the performance quality of UML models through a performance antipatterns detection and a model-based refactoring engines, which exploit Epsilon~\cite{Kolovos:2010ua} to implement detection rules and refactoring actions\footnote{PADRE: \url{https://github.com/SEALABQualityGroup/padre}}. 
Furthermore, PADRE employs UML models augmented by MARTE stereotypes in order to link performance data to UML elements. MARTE, which stands for Modeling and Analysis of Real-time and Embedded systems, is the official OMG profile that extends the UML with quantitative knowledge. The MARTE profile is structured in packages and sub-package each one with a specific aim. In our approach, we use stereotypes contained in Generic Quantitative Analysis Modeling (GQAM) package. Through the GQAM stereotypes we are able to bring the runtime data back to the UML element, and on the basis of this runtime knowledge, PADRE can detect and eventually remove performance antipatterns. A performance antipattern~\cite{Cortellessa:2014cs} is a description of well-known bad design practices that might lead to performance degradation.

Moreover, PADRE is equipped with a performance analyzer, which exploits Queueing Networks and an MVA approximation algorithm to obtain performance indices.

In particular, PADRE can detect eight performance antipatterns, and it provides several refactoring actions. A refactoring action can be either a specific action, \ie designed to remove specific antipatterns, or a general one, \ie aimed at improving the system performance quality without targeting specific aspects. PADRE provides three different detection and refactoring sessions, namely user-driven, batch and multiple sessions. The multiple sessions allows to apply more than one refactoring action in a row, the batch session performs refactoring actions until every performance antipattern has been removed, and the users-driven allows the performance expert to select a specific performance antipattern occurrence to be removed. In the presented paper, we employ the user-driver session having the performance expert part of our refactoring loop.

\subsection{Performance Antipattern}\label{sec:background:pas}

In this section we introduce the performance antipattern (PA) concept.
A performance antipattern describes a bad design practice that might lead to performance degradation in a system. This concept is mutuated from the design antipatterns one, which describes well-known bad practices that might cause system quality degradation (\eg low cohesion). 

Smith and Williams had textually described a set of performance antipatterns (PA) in~\cite{Smith:2002vl} that they have identified on the basis of existing experiences. Then, this textual descriptions have been translated in first-order logics representation~\cite{DBLP:journals/sosym/CortellessaMT14}, thus enabling the automated detection of PAs. 
The first-order logics representation of a PA is a combination of multiple literals, where each one maps on a specific system view. 
Furthermore, every literal is compared to a threshold that represents a safety limit that the system shall not overstep. 

In this paper we consider the Blob and the Pipe and Filter performance antipatterns, because they have a larger potential to occur in microservice-based systems. In the following we recap the definition of these PAs.

\paragraph{Blob} It occurs when a single component (also known as God Class) performs the most part of the work of a software system, and its manifestation results in excessive message traffic that may degrade performance. Expression~\ref{eq:pas.blob} describes the Blob performance antipattern in first-order logics~\cite{DBLP:journals/sosym/CortellessaMT14}.

\noindent The first inequality of Expression~\ref{eq:pas.blob} refers to the number of the exposed interfaces of a component, where a too high number of interfaces is considered as a precondition to identify the component as a God Class. The second inequality, instead, checks whether the God Class is effectively involved in the system. The third inequality refers to hardware utilization of hardware where the component runs. Only if all three inequalities hold then the component is identified as a Blob performance antipattern.
\begin{gather}
\nonumber \exists c_x, c_y \in \mathbb{C}, S \in \mathbb{S} \mid \\
\nonumber F_{numClientConnects}(c_x) \geq Th_{maxConnects} \quad \wedge \\
F_{numMsgs}(c_x, c_y, S) \geq Th_{maxMsgs} \quad \wedge \label{eq:pas.blob}\\ 
\nonumber F_{maxHwUtil}(P_{xy}, all) \geq Th_{maxHwUtil} \end{gather}

\paragraph{Pipe and Filter} It occurs when the slowest filter in a ``pipe'' causes the system to have unacceptable throughput. This situation is formalized in Expression~\ref{eq:pas.paf} ~\cite{DBLP:journals/sosym/CortellessaMT14}.
\begin{gather}
\nonumber \exists OpI \in \mathbb{O}, S \in \mathbb{S}, i \in \mathbb{N} \mid \\
\label{eq:pas.paf}F_{resDemand}(Op) \geq Th_{resDemand} \wedge F_{probExec}(S, OpI) = 1 \quad \wedge \\
\nonumber F_{maxHwUtil}(P_{c}, all) \geq Th_{maxHwUtil} \end{gather}

\noindent Differently to the Blob performance antipattern, Pipe and Filter identifies an operation to be the cause of performance degradation. First of all, because the operation requires a too high amount of resources to be executed (\ie a heavyweight resource demand),  as described in the first inequality of Expression~\ref{eq:pas.paf}. Beside this, the operation has to be certainly executed in order to be the cause of performance degradation, and this is checked in the second equality. Finally, either the hardware utilization must exceed a safety threshold (i.e., third inequality) for the Pipe and Filter to manifest itself.
Here the first two literals refer to design characteristics of the system, while the last two one to performance properties.

 \section{Our approach}\label{sec:approach}

\noindent
The idea underlying our approach exploits the correspondences between the architectural design and the runtime aspects of a software system, with the aim of improving its performance.

\begin{figure*}[!ht]
	\centering
	\begin{figrev}
	    \includegraphics[width=1\textwidth]{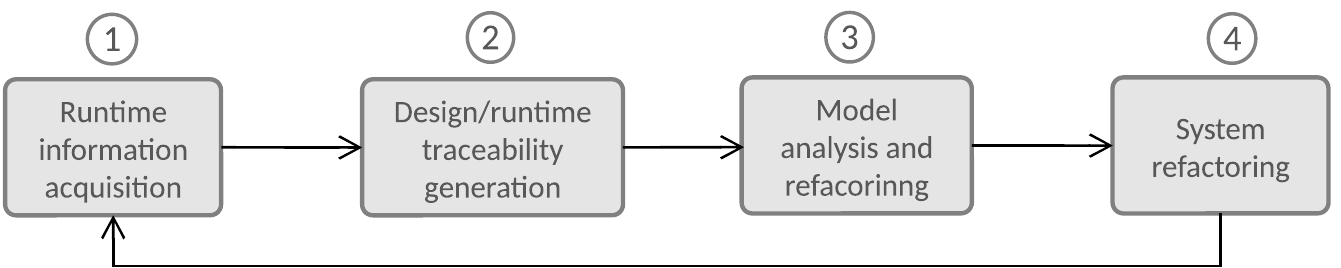}
	    \caption{A high-level workflow of our approach.}\label{fig:approach}
	\end{figrev}
\end{figure*}

Figure~\ref{fig:approach} depicts four main steps of the continuous performance engineering loop we considered, as follows:
\begin{enumerate}
	
	\item \emph{Runtime data acquisition}: Microservice-based systems are monitored by means of distributed tracing; thus, logs are stored in a central location and metrics for all instances of a given service are aggregated to understand the overall state. Then, the collected data are represented in a model-based format compliant with EMF. 

	\item \emph{Design-runtime traceability generation}: In this phase, the system behavior at runtime is matched with the architectural design by means of traceability models that are automatically generated on the base of correspondences that are formally pre-defined through a metamodel. 
	
	\item \emph{Model analysis and refactoring}: The analysis of the above created traceability models aims to connect system performance issues with the affected design components that are identified as possible causes. The results of such analysis lead to the definition of model refactoring actions, that are applied on the system model in order to identify the most promising ones.

	\item \emph{System refactoring}: The emerging refactoring actions are implemented and applied to the running system, whose runtime data is acquired and used in the next iteration of this workflow.  

\end{enumerate}

In the rest of this section, we describe each step in detail.

\subsection{Runtime data acquisition}  \label{sec:overall-runtime}
\begin{figure}[!ht]
	\centering
	\begin{figrev}
	    \includegraphics[width=1\textwidth]{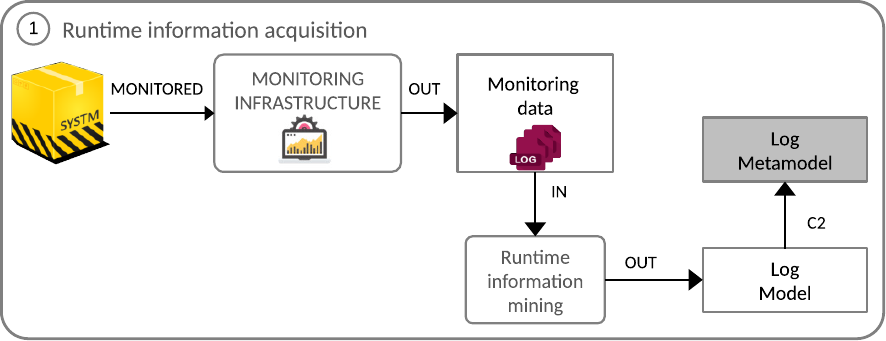}
	    \caption{Runtime data acquisition}\label{fig:approach-runtime}
	\end{figrev}
\end{figure}

\noindent
As depicted in Figure~\ref{fig:approach-runtime}, runtime data (\ie logs/traces) are obtained through a monitoring infrastructure over a running system. The specific infrastructure adopted in this work has been detailed in Section~\ref{subsec:monitoring_techniques}.

\begin{figure*}[htpb]
	\centering
	\includegraphics[width=0.8\textwidth]{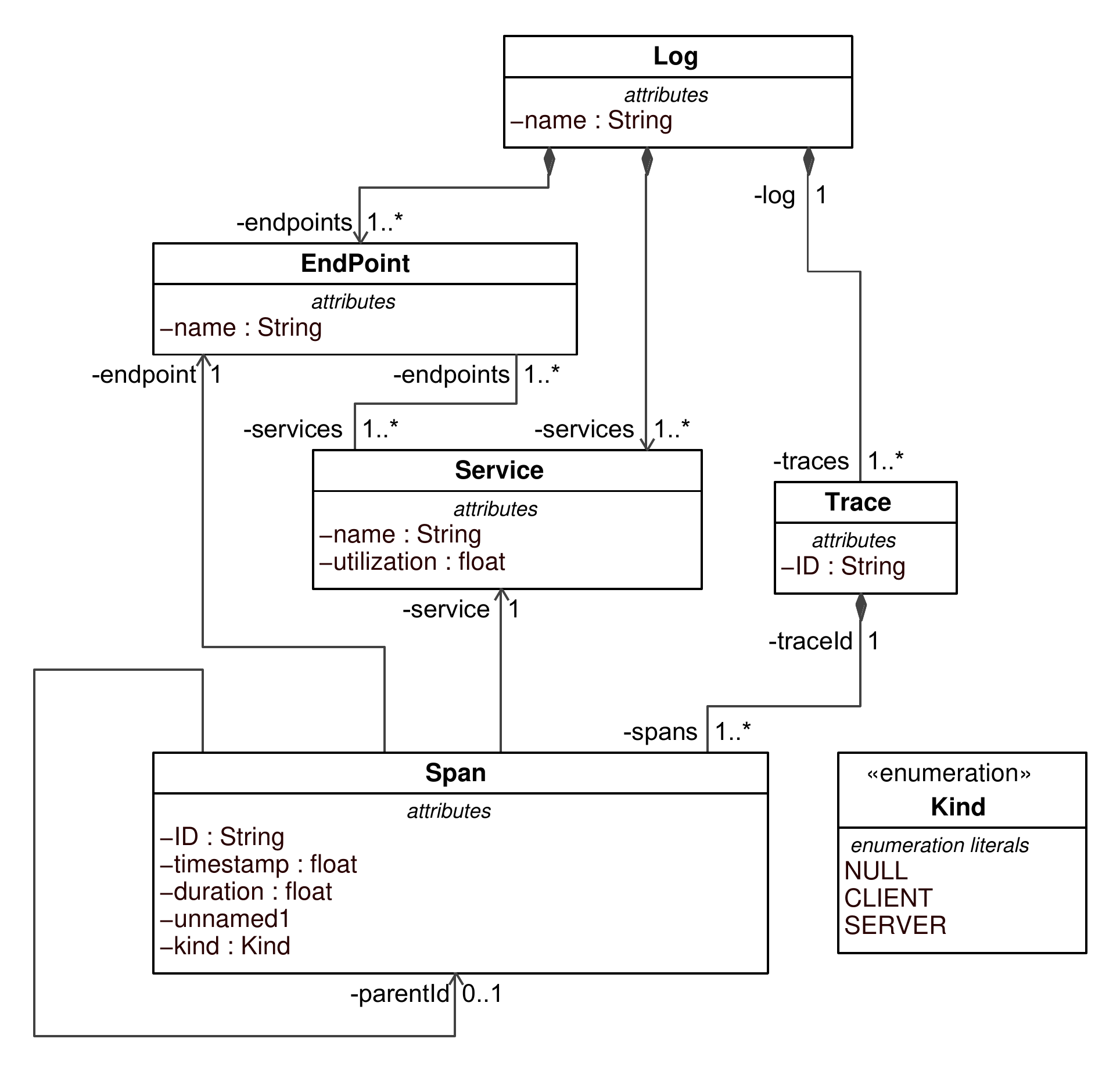}
	\caption{Log Metamodel}
	\label{fig:logmm}
\end{figure*}

\begin{figure*}[htpb]
    \centering
    \includegraphics[trim={6.5cm 0 0 1cm},clip,width=1\textwidth]{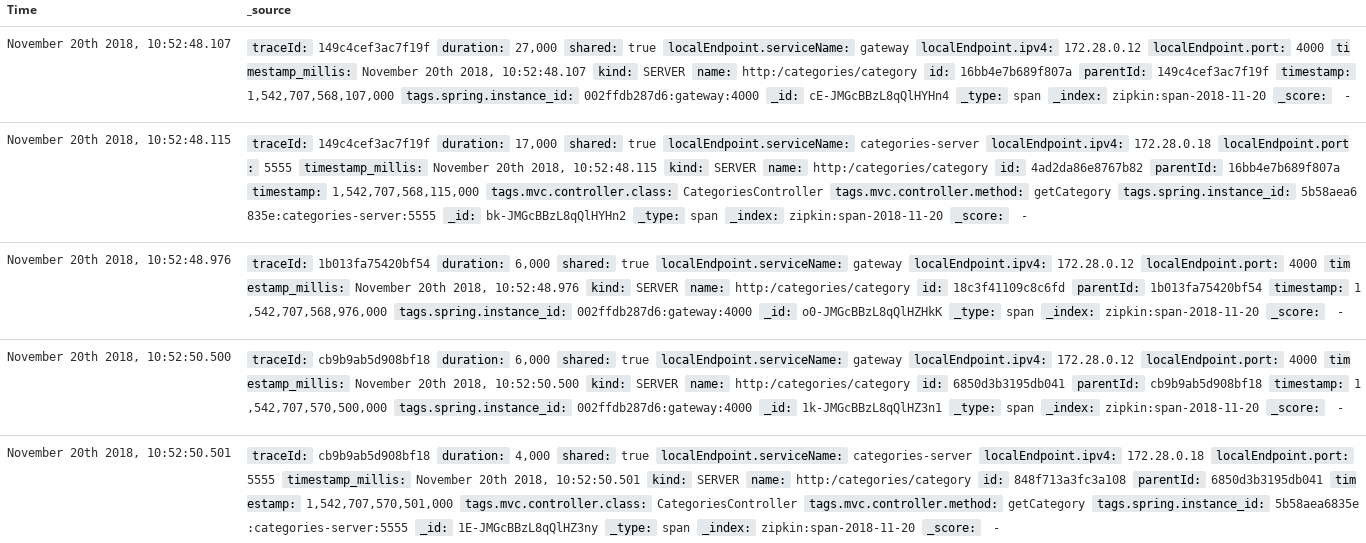}
    \caption{A fragment of the raw log} \label{fig:raw-log}
\end{figure*}

The collected runtime data is then integrated in an EMF-based environment and translated in EMF artifacts. In this step, raw logs, as the one shown in Figure~\ref{fig:raw-log}, are automatically transformed in {Log Model}s conforming to a specific {Log Metamodel} reported in Figure~\ref{fig:logmm}.

The {Log Metamodel} defines the characteristics of a {Log} element, which is the root of a log model. A Log stores all the {Trace} information about requests being sent to {EndPoint}s. {Service} elements, that may represent the microservices of an application, are associated to both {Span}s and {EndPoint}s. {Service}s also include a \emph{utilization} attribute setting the percentage of CPU usage during the observation time. 
A Trace is identified by a unique {ID} and includes a set of {Span}s representing execution events. A {Span} is defined by the following attributes: {timestamp} describing when the event occurs, \emph{duration} describing the time to complete the call, and {kind} that may be one of {SERVER}, {CLIENT} or {UNDEFINED}. Moreover, when a {Span} is triggered by another one, the {parentId} reference connects the triggered {Span} to the triggering one, called the parent {Span}. A {Span} also refers to an {EndPoint}, which is the URL used to perform a request. 

\begin{figure}[ht]
	\centering
	\includegraphics[width=0.9\textwidth]{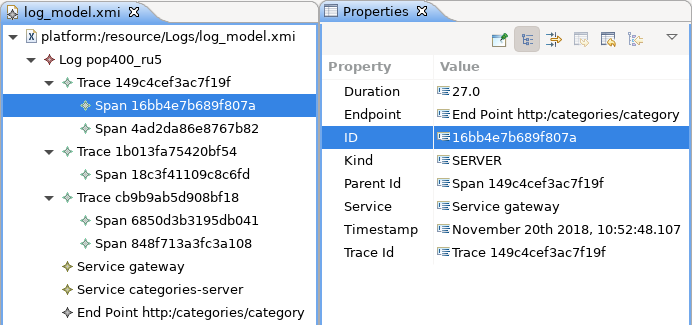}
	\caption{A Log Model sample in Eclipse}
	\label{fig:logm}
\end{figure}

Figure~\ref{fig:logm} depicts a sample of a Log Model that represents the original logs shown in Figure~\ref{fig:raw-log}, where the information that is negligible for our purposes has not been included. 
For instance, the topmost Span (id \emph{16bb4e7b689f807a}) represents the first span in Figure~\ref{fig:raw-log} with a \emph{27ms} duration, of \emph{SERVER} kind, and with the \emph{November 20th 2018 10:52:48.107} timestamp for the call to the \emph{http://categories/category} EndPoint belonging to the \emph{gateway} Service. Such a model is automatically generated from the original raw log by means of a Java transformation able to serialize the textual representation of the logs into xmi-encoded models conforming to the Log Metamodel.

\subsection{Design-runtime traceability generation} \label{sec:overall-traceability}

\begin{figure}[!htbp]
	\centering
	\begin{figrev}
	    \includegraphics[width=0.9\textwidth]{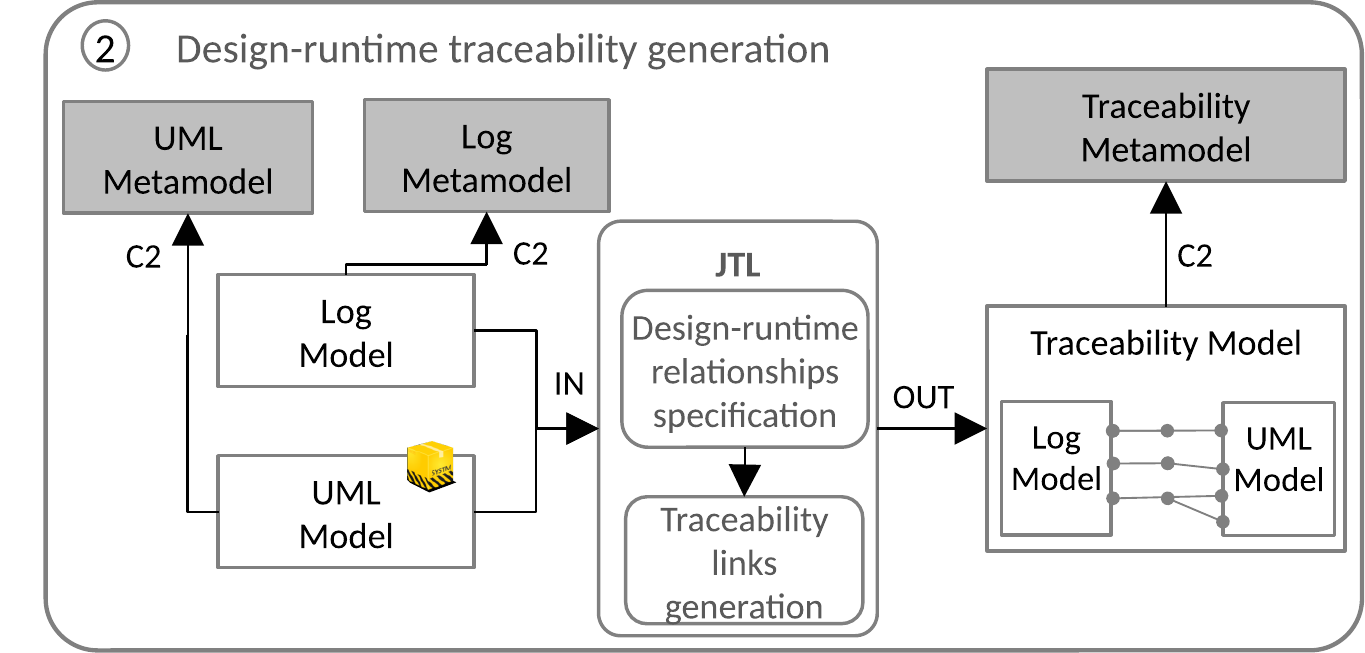}
	    \caption{Design-runtime traceability generation}\label{fig:approach-traceability}
	\end{figrev}
\end{figure}

\noindent
In this phase, as depicted in Figure~\ref{fig:approach-traceability}, the correspondences between the system behavior at runtime and the architectural design are defined and generated. 
In this work, we generate traceability links between UML and Log Models by means of JTL (introduced in Section~\ref{subsec:jtl}). In particular, JTL supports the specification of \emph{design-runtime relationships} in a declarative way at metamodel level, as bidirectional model transformations (\ie between design and log metamodels). The JTL traceability engine is able to execute such bidirectional model transformations and automatically generate the corresponding traceability links between elements of the UML design model and the log model ones.

\begin{figure}[htbp]
	\centering
	\includegraphics[width=0.7\textwidth]{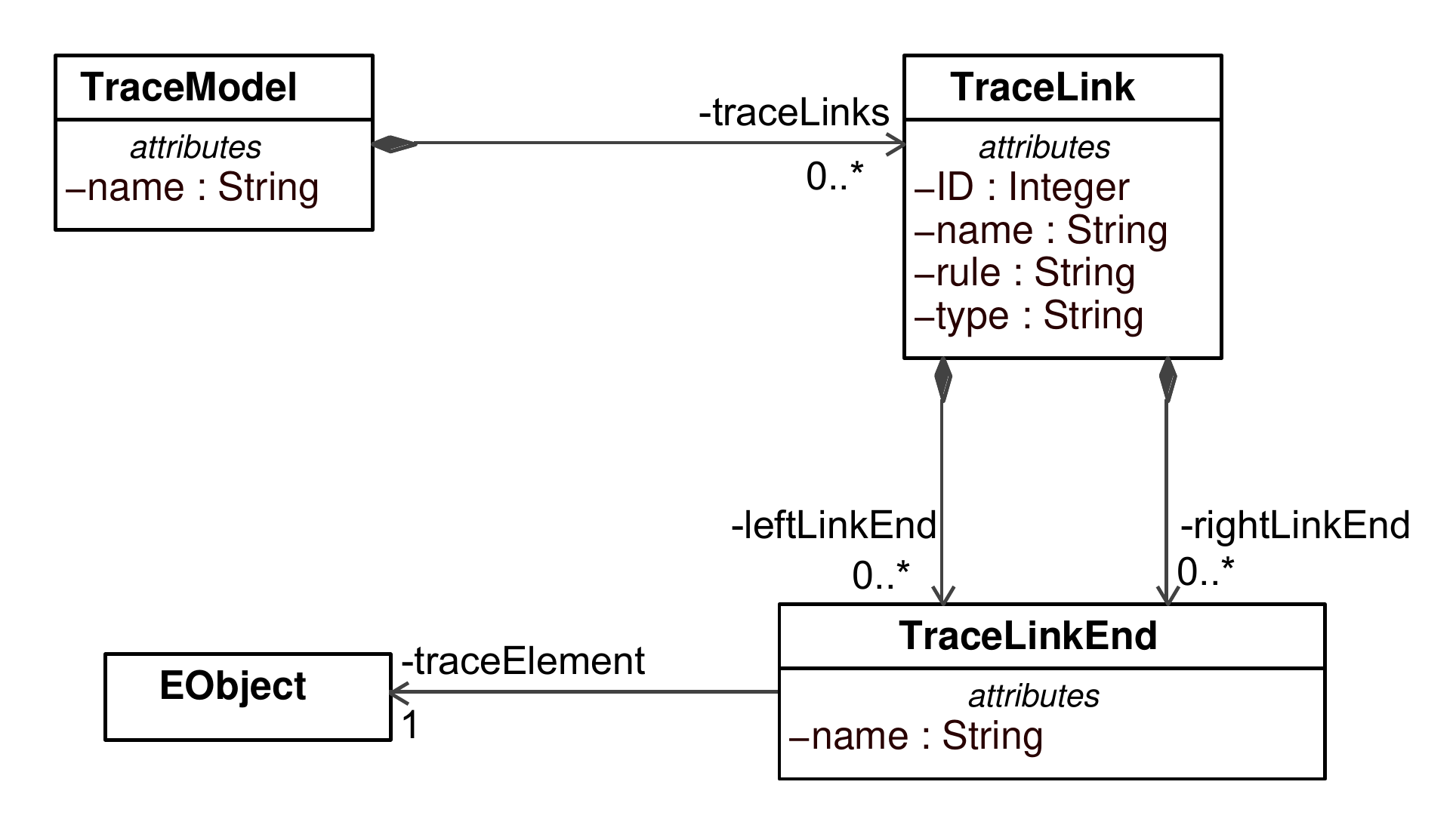}
	\caption{JTL Traceability Metamodel}
	\label{fig:tracemm}
\end{figure}

Traceability links are collected in an explicit way in Traceability Models conforming to a dedicated metamodel, namely the JTL \emph{Traceability Metamodel}. 
As depicted in Figure~\ref{fig:tracemm}, it basically defines the notion of {TraceModel},  which is the root element of a traceability model. It relates a model belonging to a "left" domain to a model belonging to a "right" domain. In particular, a set of trace links between left and right elements, as long as the rules that enforced their mapping, are collected. A {TraceLink} relates one or more elements belonging to the left domain ({leftLinkEnd}) and the corresponding (one or more) elements belonging to the right domain ({rightLinkEnd}). Such links connect elements of {TraceLinkEnd} type that have a name and a type. Each TraceLinkEnd refers to an object of {EObject} type (org.eclipse.emf.ecore.EObject) that represents a specific object in the left or right domain.

In Listing~\ref{lst:jtl} we show an excerpt of JTL-defined correspondences between the design and runtime concepts. 
While runtime concepts are represented by means of elements belonging to the Log metamodel (as explained in Section \ref{sec:overall-runtime}), software design concepts are represented by elements belonging to the UML metamodel. In particular, for behavioral aspects we target elements of Sequence Diagrams, whereas for static aspects we target Use Case, Component, and Deployment Diagrams. 

The specification is defined by means of relations between elements of the two involved metamodels. UML Use Cases are related to monitoring Traces, since they represent executions of the system. UML Messages in Sequence Diagrams are related to monitored Spans, as both represent operations occurring in a scenario. In order to map an Operation invoked by a Message to a specific API EndPoint, we relate an EndPoint of a Span to a Signature of a Message. Finally, since microservices are modeled as UML components, we relate them to Services of Spans. The above described mappings can be specified in a declarative manner as correspondences in JTL, as described in the following.

In Line~\ref{lst:transf}, variables {log} and {uml} are declared to match models conforming to the Log and UML metamodels, respectively. 
The specified relations are described as follows:
\begin{itemize}
\item[-] The top relation {Trace2UseCase} (Lines~{\ref{lst:T2UC}-\ref{lst:T2UC_end}}) maps a container element of Trace type in the Log domain, and a container element of UseCase type in the UML domain. The {where} clause invokes the execution of the {Span2Message} relation;
	\item[-]  The {Span2Message} relation (Lines~{\ref{lst:S2M}-\ref{lst:S2M_end}}) maps a Span and a Message type elements involved in a use case interaction. The {where} clause invokes the execution of the {EndPoint2Signature} relation;
	\item[-]  The {EndPoint2Signature} relation  (Lines~{\ref{lst:EP2S}-\ref{lst:EP2S_end}}) maps an EndPoint of a Span and an Operation type element that represents the signature of a message;
	\item[-]  The top relation {Service2Component} (Lines~{\ref{lst:S2C}-\ref{lst:S2C_end}}) maps a Service type container element to a Component type one. 	
\end{itemize}

\lstdefinelanguage{jtl}{
	morekeywords = {transformation,top,relation,enforce,checkonly,domain,where,when},
	morecomment=[l]{//},
	morecomment=[s]{/*}{*/}
}
\begin{lstlisting}[label={lst:jtl}, language=jtl, escapechar=|, caption={Log2UML correspondences specification}]
transformation Log2UML (log:Log, uml:UML) { |\label{lst:transf}|
  ...
  top relation Trace2UseCase {  |\label{lst:T2UC}|
    checkonly domain log t : Log::Trace {
      spans = s : Log::Span { }
    };
    checkonly domain uml uc : UML::UseCase {
      ownedBehavior = ob : UML::Interaction { 
        message = m : UML::Message { }
      }
    };
    where { Span2Message(s, m); }
  } |\label{lst:T2UC_end}|
  
  relation Span2Message { |\label{lst:S2M}|
    checkonly domain log s : Log::Span {
      endpoint = ep : Log::EndPoint { }
    };
    checkonly domain uml m : UML::Message {
      signature = s : UML::Operation { }
    };
    where { EndPoint2Signature(ep, s); }
  } |\label{lst:S2M_end}|
  
  relation EndPoint2Signature { |\label{lst:EP2S}|
    n : String;
    checkonly domain log ep : Log::EndPoint {
      name = n
    };
    checkonly domain uml s : UML::Operation {
      name = n
    };
  } |\label{lst:EP2S_end}|
  
  top relation Service2Component { |\label{lst:S2C}|
    n : String;
    checkonly domain log s : Log::Service {
      name = n
    };
    checkonly domain uml c : UML::Component {
      name = n
    };
  } |\label{lst:S2C_end}|
  ...
}
\end{lstlisting}

The described mapping assumes that the design of the system is consistent with its implementation (e.g., in terms of naming convention used). Moreover, the above described correspondences are specified according to the adopted notations. However, the approach can be extended to different modeling languages or monitoring technologies. In fact, JTL allows the specification of heterogeneous relations with different level of complexity, e.g., elements that do not trivially match by names, or relations between elements with one-to-many multiplicity~\cite{CDEP10}.

\begin{figure*}[ht]
	\centering
	\includegraphics[width=.98\textwidth]{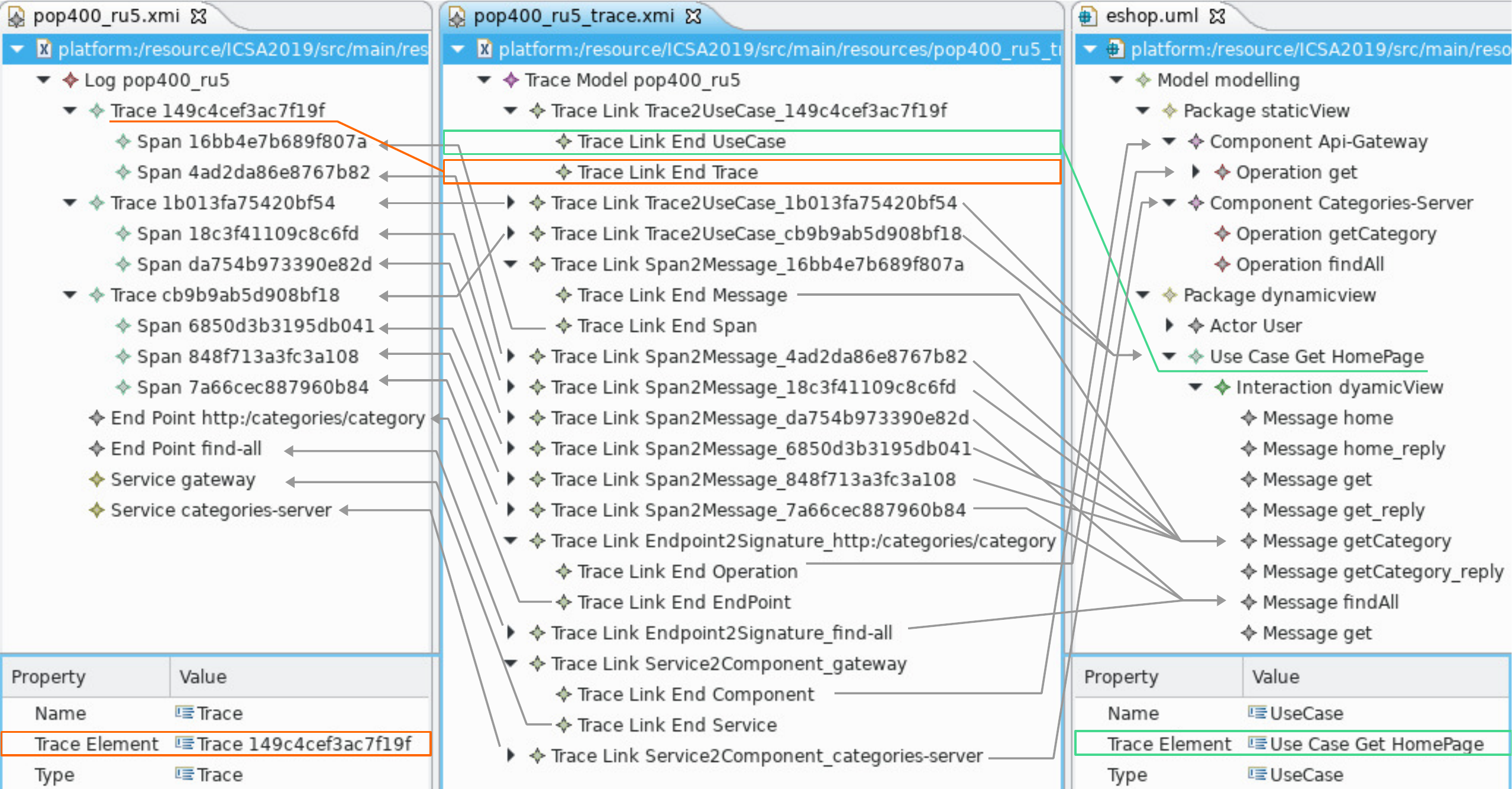}
	\caption{An example of Traceability model between Log and UML models}
	\label{fig:tracem-example}
\end{figure*}

The application of the \emph{Log2UML} transformation on a pair of Log and UML models, as shown in the left and right part of Figure~\ref{fig:tracem-example}, generates the corresponding Traceability model in the middle part of the figure. In particular, the arrows in the figure cross trace links that connect the source and target model elements they refer to.

For instance, the \emph{Trace2UseCase\_149c4cef3ac7f19f} traceability link relates the \emph{Get HomePage} use case in the right end and the corresponding \emph{149\-c4c\-ef3\-ac7\-f19f} log trace in the left end. Hence, for each message in the use case, we are able to know when the corresponding operation has started and its response time. As a consequence, the traceability model can be used to map complex performance measures, such as the average response time of a specific scenario or the average service time of an operation. This process will be described in detail in the next section. 
 
\subsection{Model analysis and refactoring} \label{sec:overall-performance}

\begin{figure}[!ht]
	\centering
	\begin{figrev}
	    \includegraphics[width=1\textwidth]{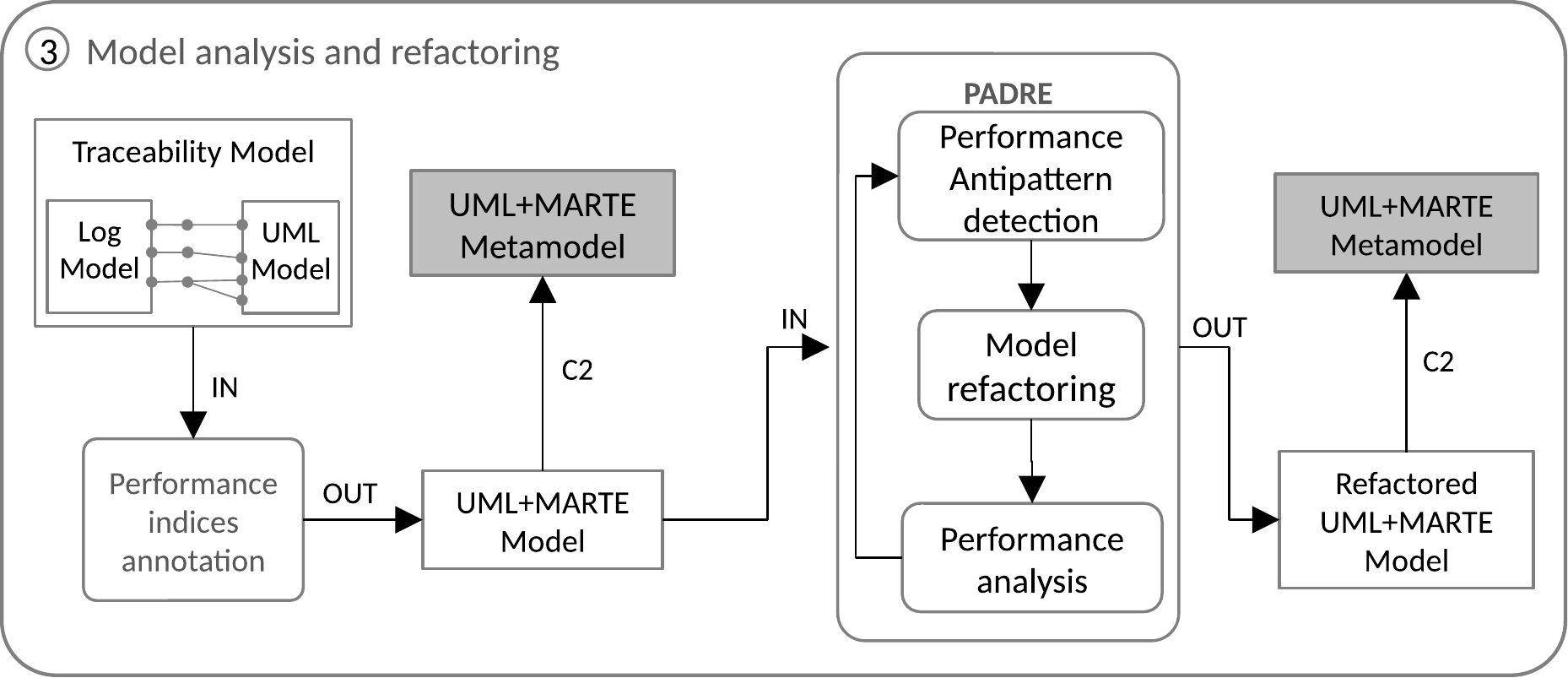}
	    \caption{Model analysis and refactoring}\label{fig:approach-detection}
    \end{figrev}
\end{figure}

\noindent

This step is aimed at analyzing the design model and removing possible performance flaws. In particular, we use runtime data (\ie traces) to augment the design model, and then we execute a performance analysis driven by antipatterns~\cite{Cortellessa2011} on the augmented model. 

The \emph{Performance indices annotator} in Figure~\ref{fig:approach-detection} exploits the Traceability Model to report runtime data back to the model, and for this goal it exploits the MARTE stereotypes, in that they allow: i) to set input data (\ie performance parameters) of performance analysis (\eg operations service demand), and ii) to fill the output data (\ie performance indices) of a performance analysis back to the model (\eg utilization).

We adopt the  \emph{MARTE:GAQM} package stereotypes, among MARTE ones, as follows:

\begin{itemize}
	\item \emph{Input} Data:
	\begin{itemize}
			\item \emph{GaWorkloadEvent:generator}: it expresses the generational value of a workload. For example, we annotate here the exponential arrival rate  $\lambda$ for an open class of jobs.
			\item \emph{GaWorkloadEvent:pattern}: it denotes if the job class is open or closed;
			\item \emph{GaAcqStep:servCount}: it expresses the service demand of a UML Operation. Hence, we annotate the UML Message that trigger that Operation in a UML Sequence Diagram to express its demand in that scenario;
	\end{itemize}
	\item \emph{Output} Data:
		\begin{itemize}
			\item \emph{GaScenario:respT}: it expresses a response time. We use it on a UML Use Case to report the response time of that scenario under a specific workload.
			\item \emph{GaScenario:throughput}: it expresses a throughput. We use it on a UML Use Case similarly to a response time;
			\item \emph{GaExecHost:utilization}: it expresses an utilization. We use it for a UML Node within a Deployment Diagram.
		\end{itemize}
\end{itemize}

In order to collect operation service demands, we stimulate the system with a lightweight workload. Indeed, in this parameterization step we aim to avoid generating queues in the system, so that the \emph{Service Demand} $D = V * S $ definition holds~\cite{Lazowska:1984ta}. In particular, $D$ is the Service Demand, $V$ is the number of visits, and $S$ is the service time. Following the definition, the service demand of an operation, in our case, is given by its response time when the lightweight workload is executed, because we guarantee (by observation) that waiting time in queue is never originated by that workload.

Once \emph{Service Demand}s have been collected and filled back to the input \emph{GaAcqStep:servCount} tags, we stimulate the system with a selected (possibly heavy and realistic) workload to discover performance flaws.

\begin{figure}[ht]
	\subfloat[Static view (Component Diagram)]{
		\includegraphics[width=0.45\textwidth]{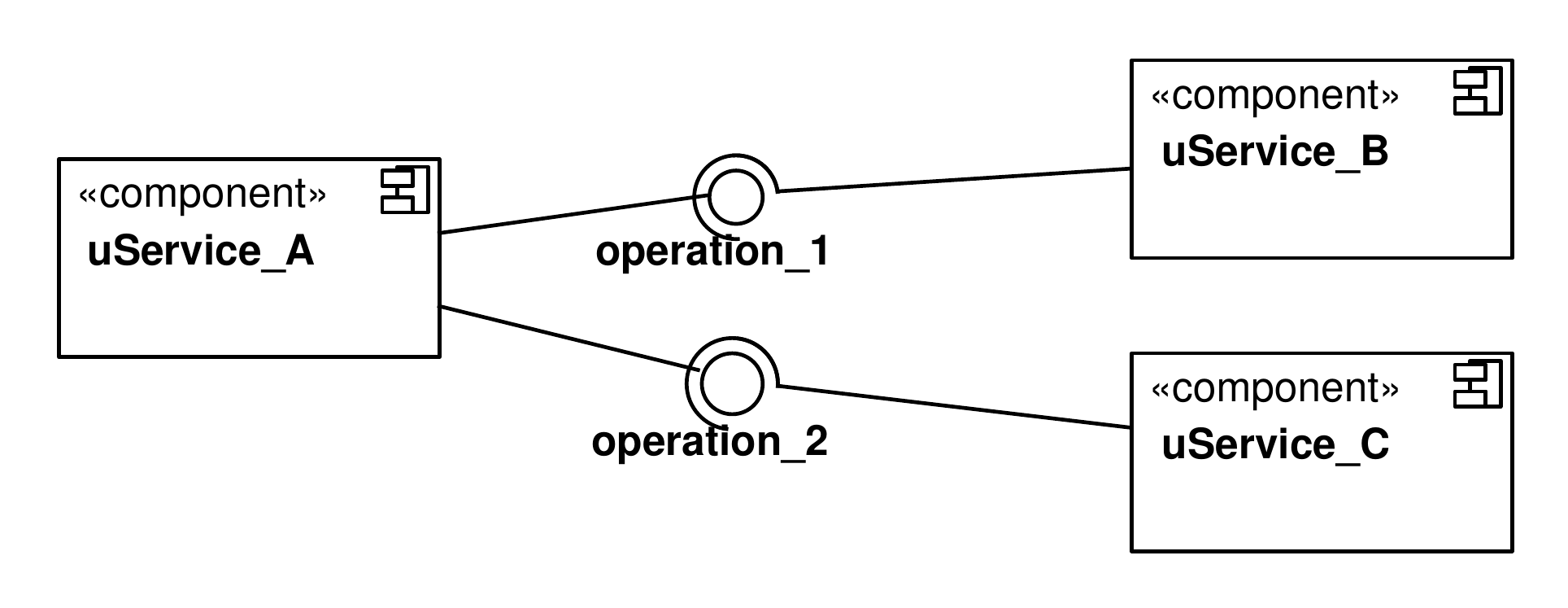}\label{fig:example-comp}}
	\subfloat[Deployment view (Deployment Diagram)]{
		\includegraphics[width=0.49\textwidth]{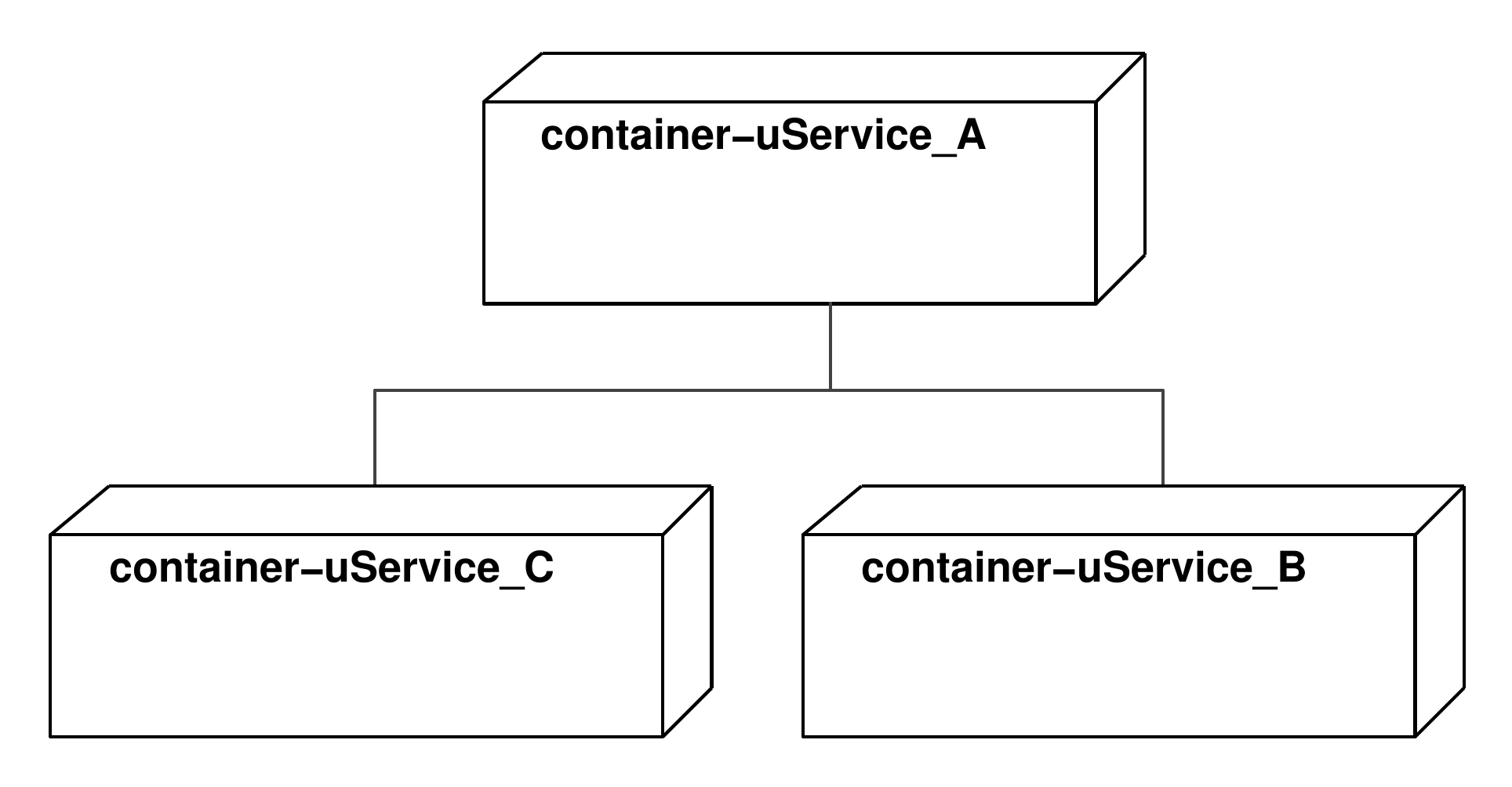}\label{fig:example-deploy}}
	\\
	\centering
	\subfloat[Static view (Use Case Diagram)]{
		\includegraphics[width=0.45\textwidth]{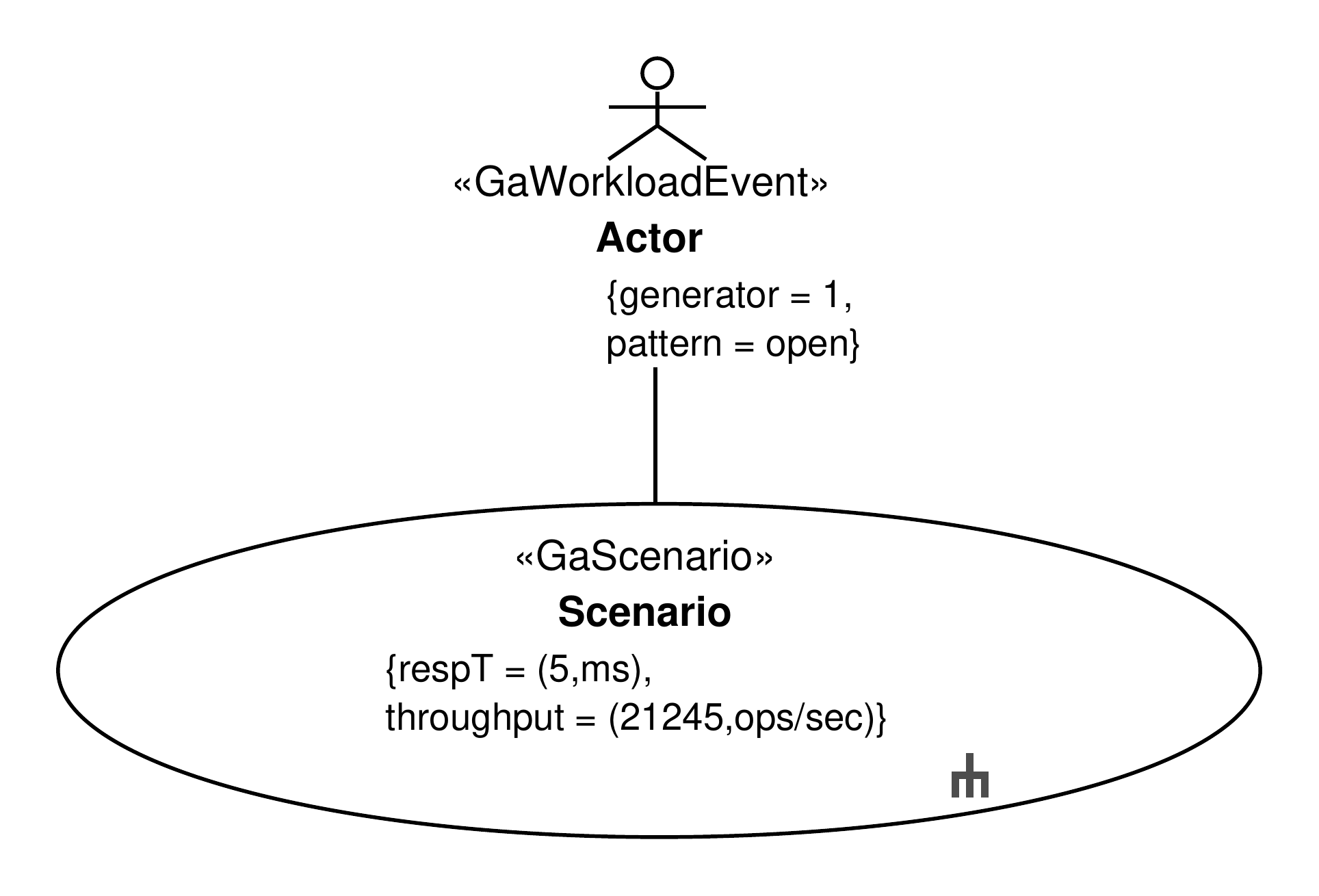}\label{fig:example-uc}}
	\subfloat[Dynamic view (Sequence Diagram)]{
		\includegraphics[width=0.45\textwidth]{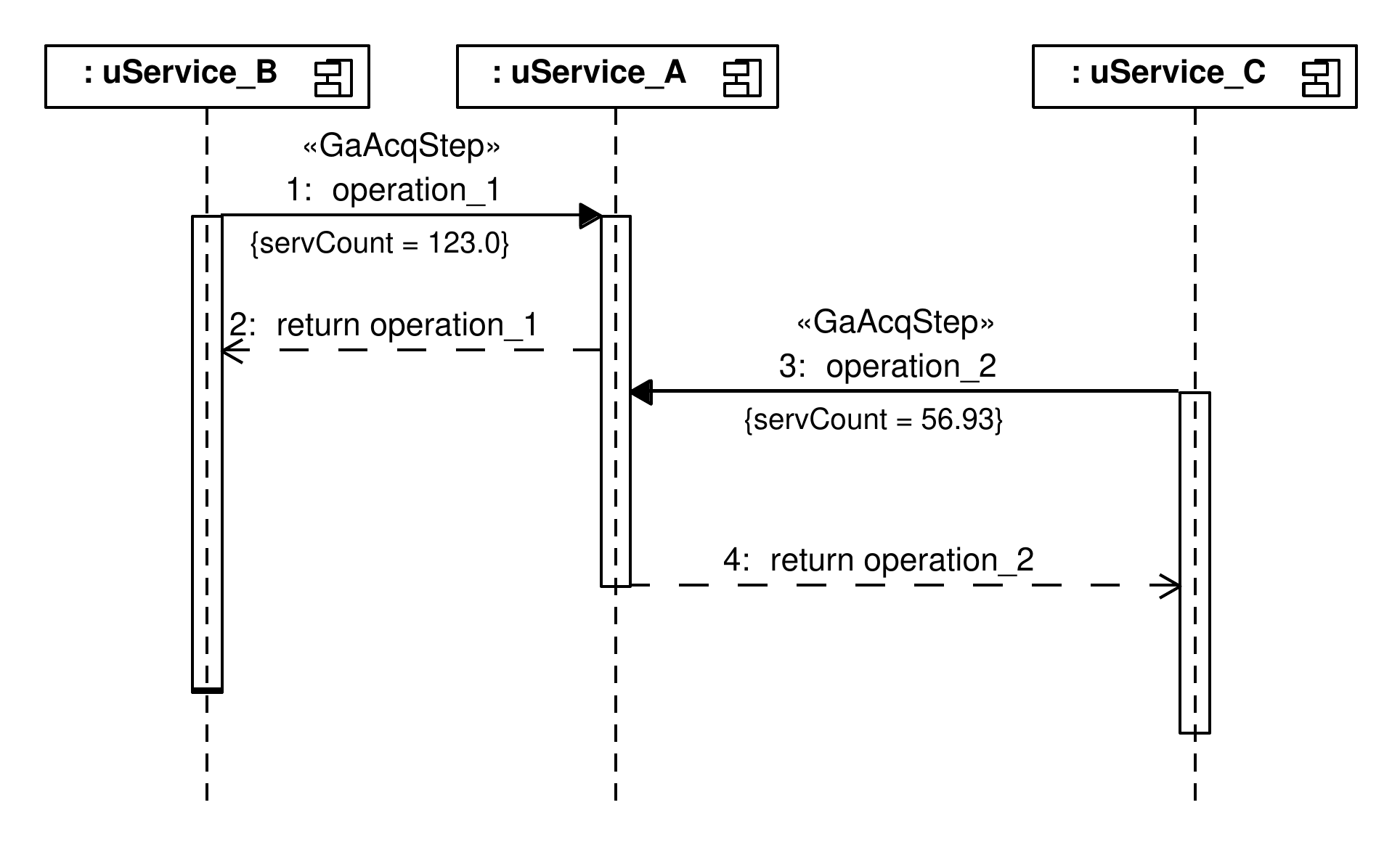}\label{fig:example-seq}}
	\caption{The \textit{example} UML Software Model}
	\label{fig:example-diagrams}
\end{figure}

Upon the system execution is completed under the selected workload, the \emph{Performance Indices Annotator} fills the performance indices tags back to the model. Then, a performance-driven refactoring loop can start. PADRE \cite{ARCELLI2018366} has been adopted for this goal, as it is an approach that detects performance antipatterns and provides a list of possible refactoring actions that shall mitigate the performance degradation. 

The PADRE refactoring loop is made of three main steps: i) \emph{Performance Antipattern Detection}; ii) \emph{Model Refactoring}; iii) \emph{Performance Analysis}. First, the \emph{Performance Antipattern Detection} is executed in order to detect performance antipatterns occurrences. It is worth noticing that PADRE employs multi-views models to discover performance antipatterns. For this reason we use a multi-view UML design model, as depicted in Figure~\ref{fig:example-diagrams}. In case performance antipatterns arise in the model, a \emph{Model Refactoring} step is performed in order to remove them. In this step, PADRE provides a list of possible refactoring actions for each performance antipattern.\footnote{The complete PADRE refactoring action portfolio is described in~\cite{ARCELLI2018366}} While executing one refactoring action at a time, the refactored design model is given as input to the PADRE \emph{Performance analysis} step, in which a model transformation is executed to transform the UML-MARTE design model into a closed Queueing Model~\cite{Cortellessa2011}.\footnote{This kind of transformation is out of scope of this paper, thus we do not provide here more details.} Thereafter, the Queueing Model is solved through the \emph{Mean-Value Analysis (MVA)} algorithm~\cite{DBLP:journals/jacm/ReiserL81}, which allows to rapidly carry out performance indices. The latter ones are exploited to recognize whether the refactoring action is promising or not.

In this paper, we have restricted the PADRE refactoring actions portfolio to the actions that we found more appropriate for a microservice context, namely:

\begin{itemize}
	\item  Clone refactoring:

	The clone refactoring action is aimed at introducing a replica of a microservice. In our modeling assumptions, we consider a microservice as a \emph{UML Component}, and a docker container as a \emph{UML Node}. The action at a glance is shown in Figure~\ref{fig:ref-clone-uml-diagrams}. 
	
	Figure~\ref{fig:clone-comp} and~\ref{fig:clone-deploy} depict the initial design model, while Figure~\ref{fig:ref-clone-comp}, and~\ref{fig:ref-clone-deploy} show the refactored design model. 
	The clone refactoring action in a nutshell: i) creates a new UML Component (\ie \emph{cloned-uService\_A}), and ii) creates a new UML Node (\ie \emph{cloned-container-uService\_A}) on which the replica is deployed. 
	In this particular case, the dynamic view is not depicted because the refactoring action does not affect it.
		
	\begin{figure}[ht]
		\subfloat[Initial]{
			\includegraphics[width=0.46\textwidth]{figures/staticview.pdf}\label{fig:clone-comp}}
		\subfloat[Refactored]{
			\includegraphics[width=0.46\textwidth]{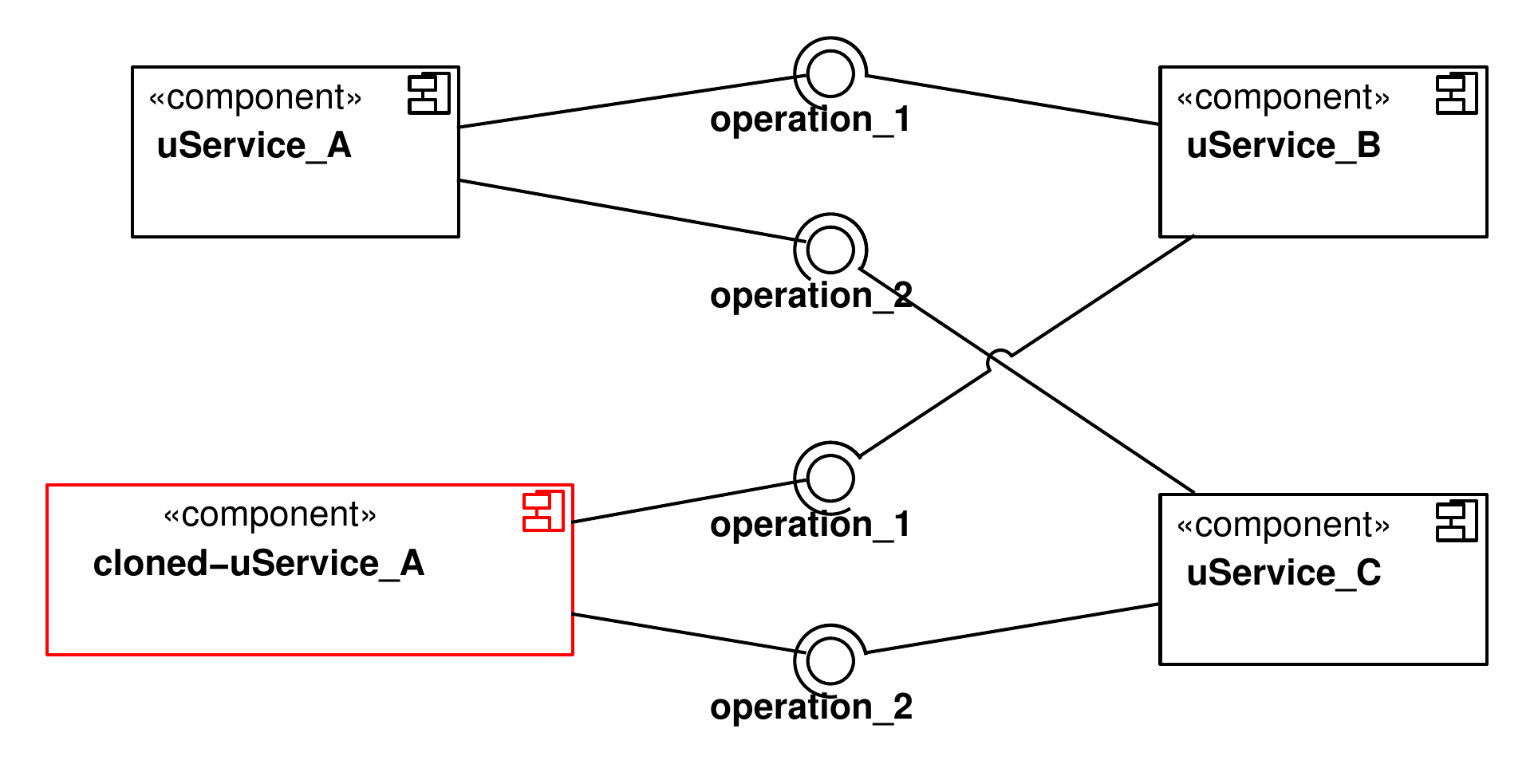}\label{fig:ref-clone-comp}}
		\\
		\subfloat[Initial]{
			\includegraphics[width=0.46\textwidth]{figures/deployment.pdf}\label{fig:clone-deploy}}
		\subfloat[Refactored]{
			\includegraphics[width=0.46\textwidth]{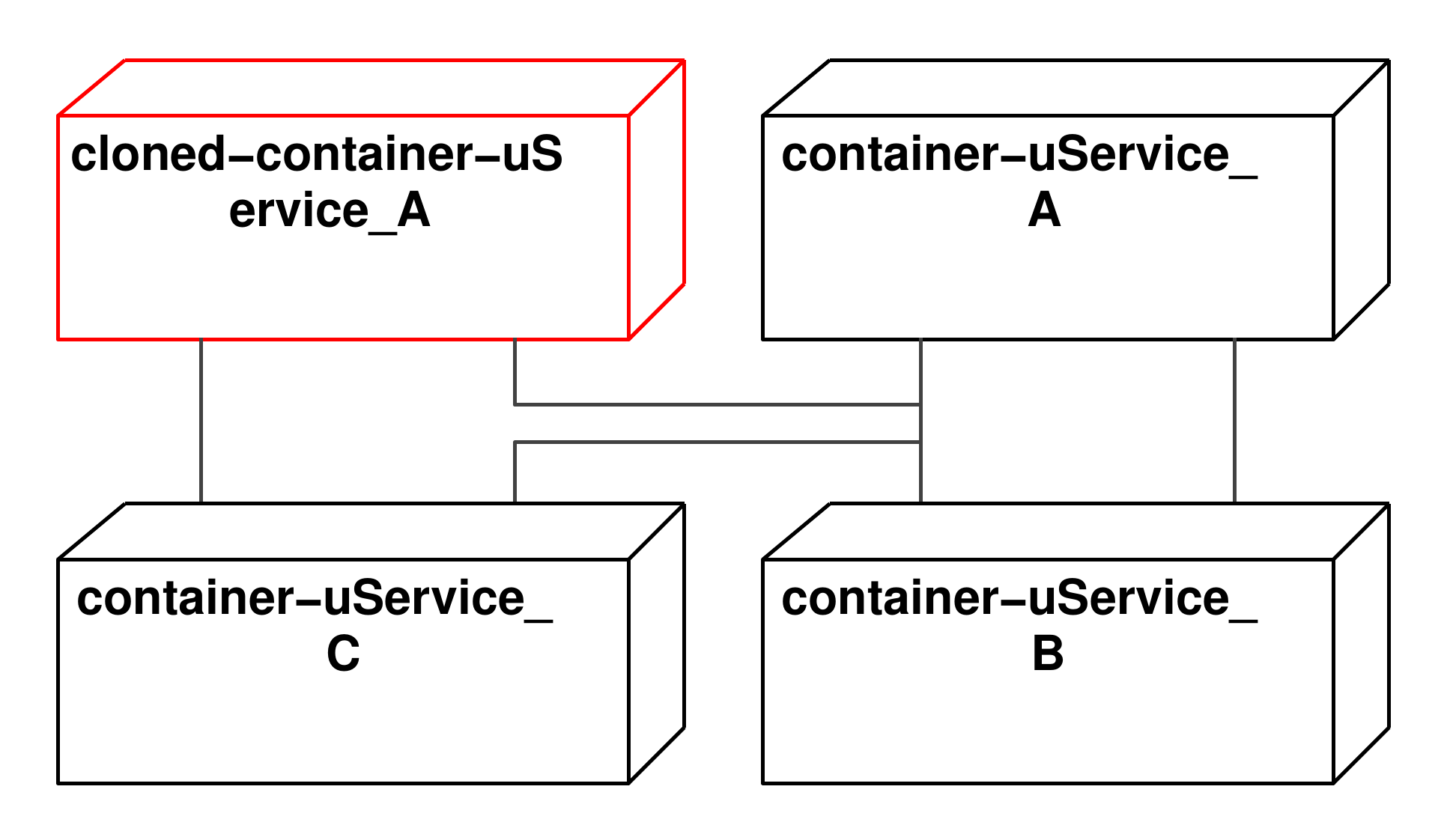}\label{fig:ref-clone-deploy}}
		\caption{The \textit{clone}  refactoring action example on component \emph{uService\_A} through a UML Software Model.}
		\label{fig:ref-clone-uml-diagrams}
	\end{figure}

	\item  Move operation refactoring:
	
	The move operation refactoring action is aimed at moving a ``critical'' operation (\eg due to its Service Demand) to a new microservice.  Figures~\ref{fig:moveop-comp},~\ref{fig:moveop-dynamic} and~\ref{fig:moveop-deploy} depict the initial design model example, while Figures~\ref{fig:ref-moveop-comp},~\ref{fig:ref-moveop-dynamic} and~\ref{fig:ref-moveop-deploy} show the refactored version. In the example, we move \emph{operation\_2} of \emph{uService\_A} microservice. Thus, the action creates a replica of this microservice (\ie \emph{cloned-uService\_A}) and then it changes the behavior and deployment views, respectively.  
	
	Differently to the Clone refactoring action, the move operation involves the dynamic view. Therefore, a new UML Lifeline (\ie \emph{cloned-uService\_A}) representing the replicated microservice is created (see Figure~\ref{fig:ref-moveop-dynamic}). Then, every message referring to operation\_2 is now transferred towards the new lifeline. Furthermore, the Deployment Diagram is refactored as well. A new UML Node (\ie \emph{cloned-container-Service\_A}) is created and, finally, the action defines the newly required connections among UML Nodes (\ie the connection between the UML Node \emph{cloned-container-uService\_A} and \emph{container-uService\_C}). In particular, the new node is linked to all other nodes that were originally connected to the node on which the microservice hosting the ``critical'' operation is deployed.
	
	\begin{figure}[ht]
		\subfloat[Initial]{
			\includegraphics[width=0.46\textwidth]{figures/staticview.pdf}\label{fig:moveop-comp}}
		\subfloat[Refactored]{
			\includegraphics[width=0.46\textwidth]{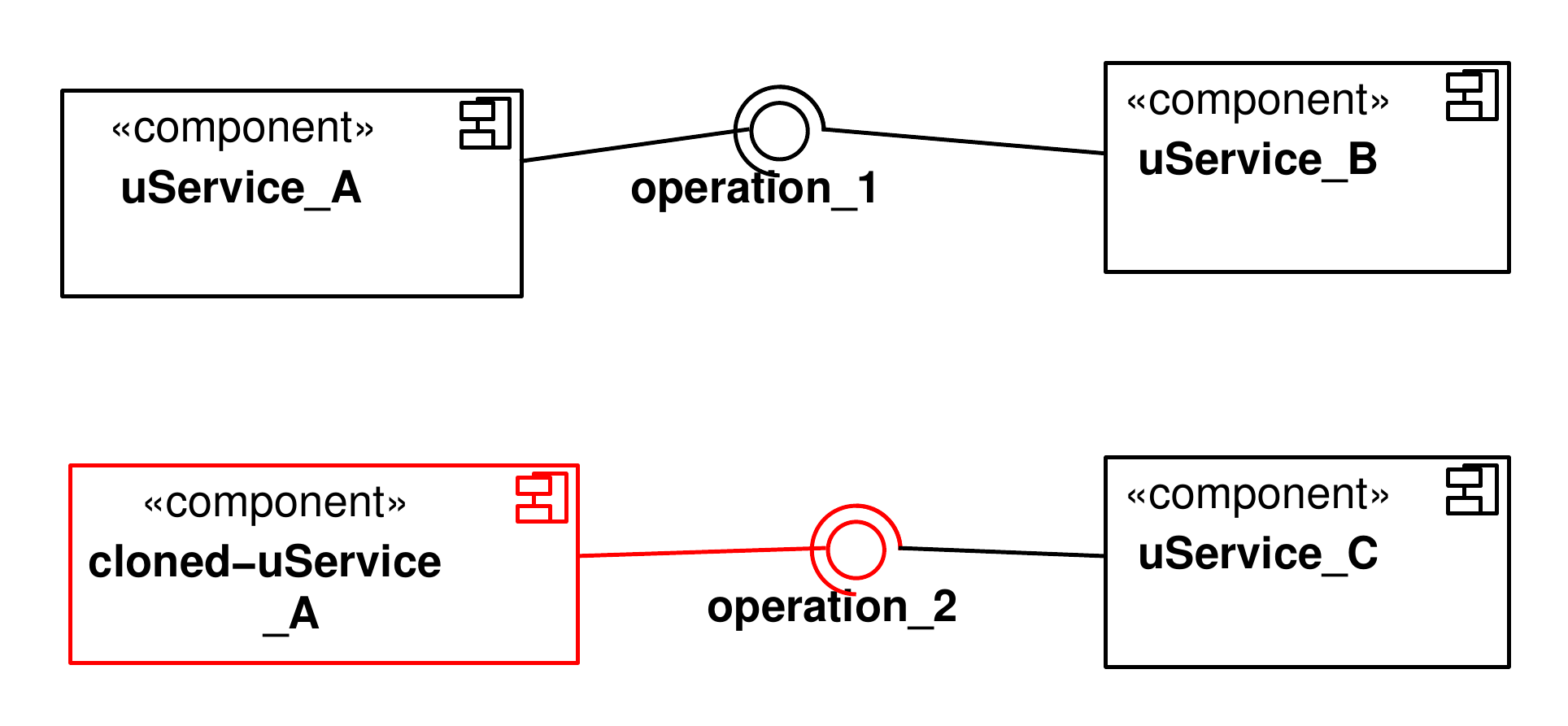}\label{fig:ref-moveop-comp}}
		\\
		\subfloat[Initial]{
			\includegraphics[width=0.46\textwidth]{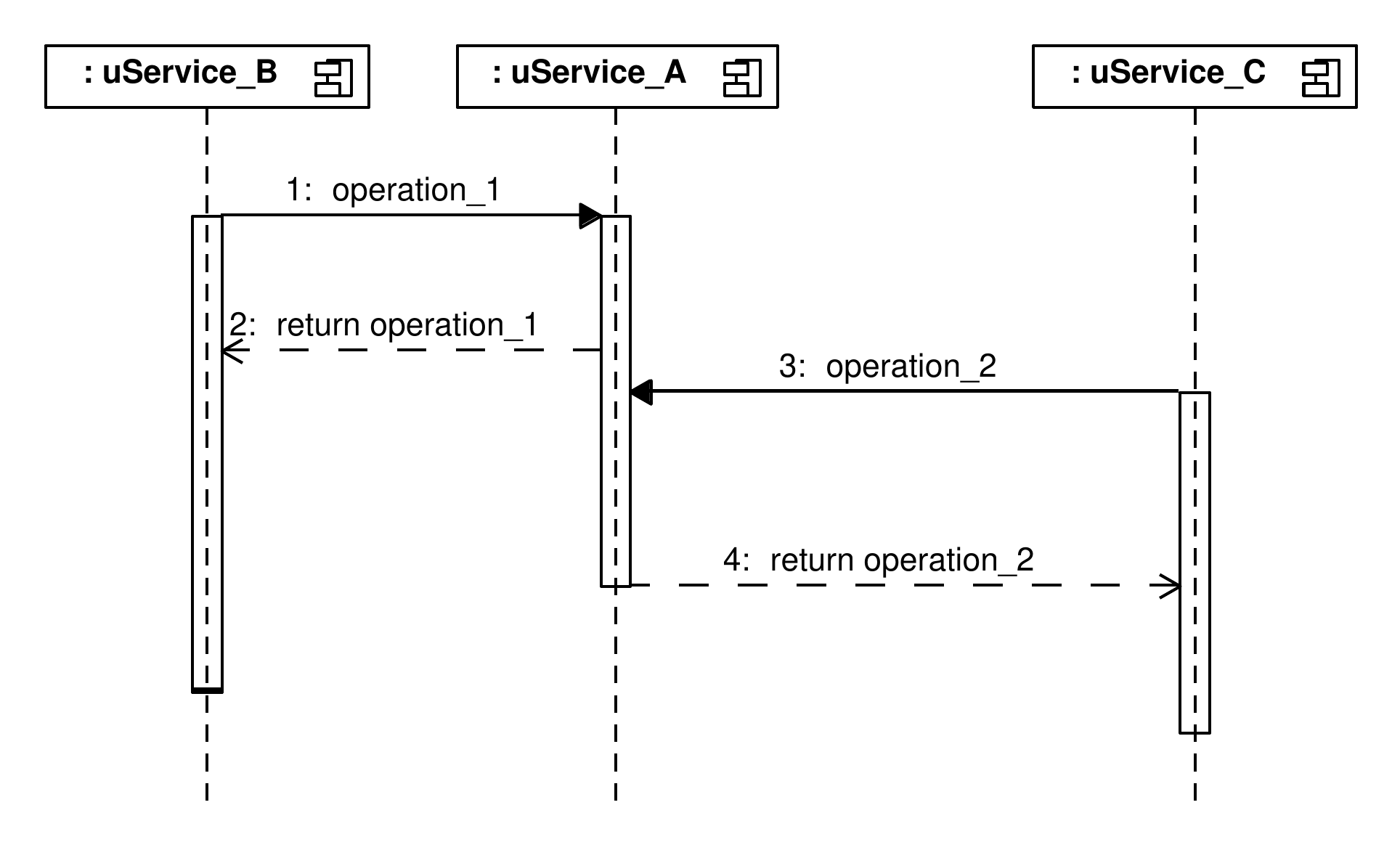}\label{fig:moveop-dynamic}}
		\subfloat[Refactored]{
			\includegraphics[width=0.46\textwidth]{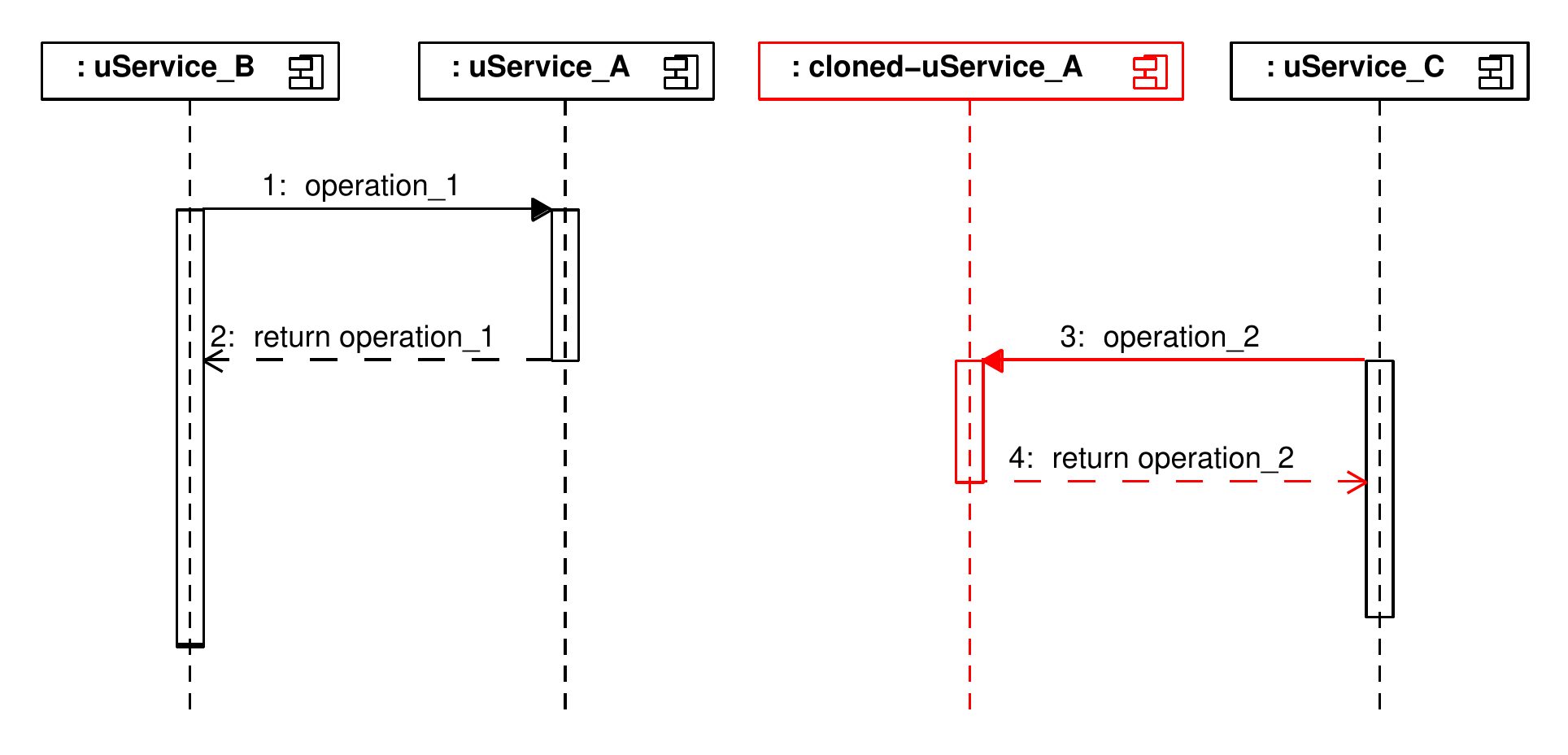}\label{fig:ref-moveop-dynamic}}
		\\
		\subfloat[Initial]{
			\includegraphics[width=0.46\textwidth]{figures/deployment.pdf}\label{fig:moveop-deploy}}
		\subfloat[Refactored]{
			\includegraphics[width=0.46\textwidth]{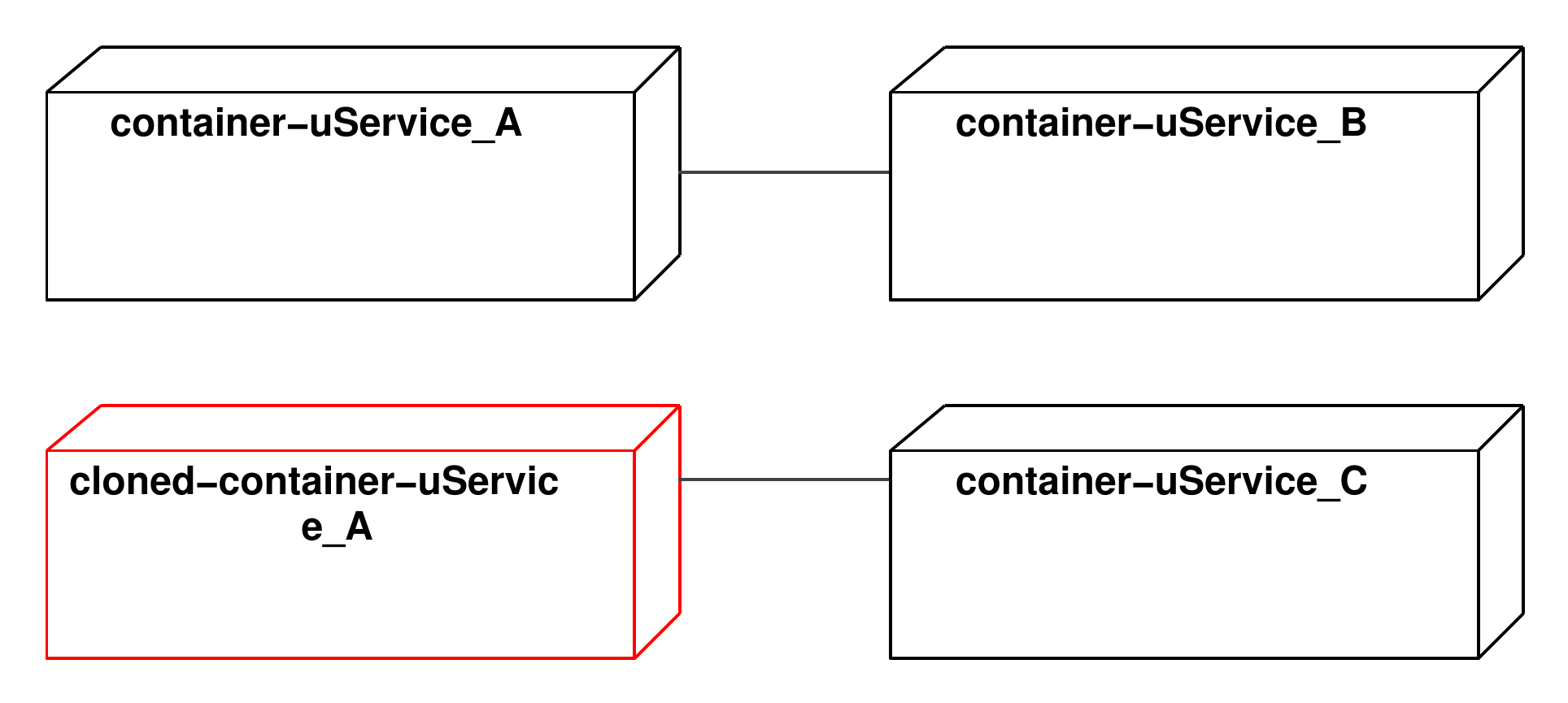}\label{fig:ref-moveop-deploy}}
		\caption{The \textit{move operation} refactoring action example on \emph{operation\_2} through a UML Software Model}
		\label{fig:ref-moveop-uml-diagrams}
	\end{figure}

\end{itemize} 
 
\subsection{System Refactoring} \label{sec:overall-refactoring}

\begin{figure}[!ht]
	\centering
	\begin{figrev}
	    \includegraphics[width=.8\textwidth]{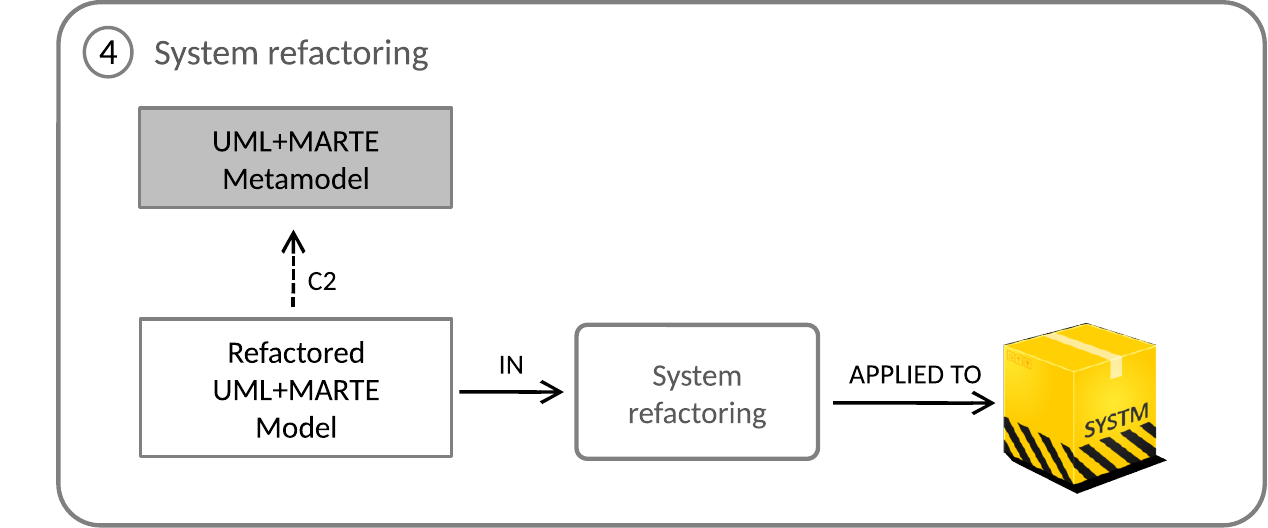}
	    \caption{System refactoring}\label{fig:approach-refactoring}
	\end{figrev}
\end{figure}

\noindent

In this step, the refactoring actions performed on the model are translated into changes of a microservice-based system.

Refactoring actions on the system have been implemented using the \emph{Docker Client}\footnote{Docker Client: \url{https://github.com/spotify/docker-client}} Java library for the operations performed on docker instances. Regarding the online modifications of configuration files, we developed a Java library that is publicly available.\footnote{Microservices refactoring library: \url{https://git.io/JLEJZ}}

\paragraph{Clone refactoring}
This action creates a replica of a microservice. In the running system, this translates into the creation of a new container deploying the same \emph{Spring Boot} microservice we intend to clone. This refactoring is achieved by exploiting the Docker API to create and start a new container using the image of the original microservice. Once the replica is up and running, we need to balance it along with the original microservice. Specifically, we want to ensure that half of the traffic that was targeted at the original microservice is now redirected to its clone. Depending on the technology used to forward requests among microservices, three different scenarios are open:
\begin{itemize}

    \item \emph{Zuul and Eureka}. This scenario is straightforward because the combination of the \emph{Zuul}\footnote{Zuul: \url{https://github.com/Netflix/zuul}} proxy with the \emph{Eureka}\footnote{Eureka: \url{https://github.com/Netflix/eureka}} registration service automatically balances the additional microservice. When the new microservice registers to \emph{Eureka}, \emph{Zuul} adds it to the physical locations available for requests forwarding. Internally, \emph{Zuul} uses \emph{Ribbon}\footnote{Ribbon: \url{https://github.com/Netflix/ribbon}} to balance incoming requests using a round-robin policy. Therefore, no further modifications are required in this case.

    \item \emph{Nginx}. When \emph{Nginx}\footnote{Nginx: \url{https://nginx.org/}} is used as a reverse proxy, we need to modify its configuration to add a new server group containing the original and the cloned microservices. Moreover, the mapping of requests has to be updated to address the requests to the newly created server group. By default, Nginx uses a round-robin algorithm to balance the servers in a server group.

    \item \emph{No proxy}. Finally, when no proxy is deployed in front of the original microservice, we can add one without disrupting the running system. In this case, we preferred to deploy \emph{HAproxy}\footnote{HAproxy: \url{http://www.haproxy.org/}}, because it is able to automatically generate a configuration by deriving the composition of server groups from the network links among docker instances.

\end{itemize}

\paragraph{Move operation refactoring}
This action moves an operation from a service to a newly created one that has the purpose of exclusively offering the operation. This is implemented by creating a replica of the original service that contains the operation we want to move and, consequently, by forwarding all the requests that were intended for the moved operation to the replica.
Similarly to the clone refactoring action, this action can be implemented in the same three scenarios:
\begin{itemize}

    \item \emph{Zuul and Eureka}. Once the replica has been created and has registered itself to \emph{Eureka}, \emph{Zuul} automatically detects it. In order to forward the requests for the moved operation to the replica, we need to add a new route in the \emph{Zuul} configuration file. Such route is needed to map the path of the moved operation to the endpoint URL of the replica.

    \item \emph{Nginx}. Analogously to the previous scenario, also when using \emph{Nginx} as a reverse proxy, we need to add a route to redirect the requests to the moved operation. This is accomplished in \emph{Nginx} by using the \emph{location} directive to map a path to an endpoint URL.

   \item{No proxy}. In this case, a new proxy is added to the application. Both \emph{HAproxy} and \emph{Nginx} can serve this purpose with similar configurations for redirecting URL paths.

\end{itemize}
  
\section{Evaluation}\label{sec:evaluation}

\noindent
In this section, we discuss the evaluation we have performed with the aim of answering the following research questions: 
\begin{itemize}
\item \emph{RQ1: Do the proposed model refactoring actions improve the performance of the running system?}
\item \emph{RQ2: To what extent does performance antipattern (PA) removal improve the whole performance?}
\end{itemize}

\subsection{Case studies setup}\label{sec:evaluation:setup}

\noindent
In order to validate the approach, we considered the following case studies: 
\begin{itemize}
	\item \emph{E-Shopper}\footnote{E-Shopper: \url{https://github.com/SEALABQualityGroup/E-Shopper}} is an e-commerce web application. The application is developed as a suite of small services, each running in its own Docker container and communicating with RESTful HTTP API. E-Shopper is composed by 9 application microservices developed with the Spring framework, each requiring a different database to operate.
	\item \emph{TrainTicket}\footnote{Train Ticket: \url{https://github.com/SEALABQualityGroup/train-ticket}} is a web ticketing application within the railway domain. It has been developed by Zhou \etal and it has been also presented in~\cite{DBLP:conf/kbse/ZhouPX0LJD18,DBLP:conf/icse/ZhouPX0XJZ18,DBLP:conf/sigsoft/Zhou0X0JLXH19}. It is made up of 40 microservices and uses different programming languages. The most used framework is again Spring, since the most used programming languange in Train Ticket is Java, and for this reason we have selected it for our evaluation. In our previous work we reverse engineered its UML representation and presented it as a reference case study~\cite{DBLP:conf/staf/Pompeo0CE19}, while here we employ it to test our performance improvement approach.
\end{itemize}

The UML models~\cite{UML2} of both case studies are augmented with  the MARTE profile stereotypes~\cite{MARTE}. We adopt MARTE because it is the standard profile widely adopted to annotate UML models with real-time and performance attributes, such as \emph{response time values}. In particular, we adopt the \emph{Generic Quantitative Analysis Model (GQAM)} package, which has been introduced to support accurate and trustworthy evaluations based on formal quantitative analyses.  

We have generated different scenarios for each application to stimulate different parts of the system and discover which ones may suffer from performance degradation under concurrent usage.

In particular, we have identified the scenarios that are more typically triggered in these applications, thus to validate our approach in performance-critical contexts. We have observed these scenarios running under different workloads, and then we have picked the workloads that stress the application performance, but that at the same time do not lead to fully system saturation, because our approach is intended to work before extreme degradation of performance occur.

For the E-Shopper application we have realized three scenarios with specific workloads:
\begin{itemize}[topsep=0pt,itemsep=0pt]
    \item \emph{Desktop}, the request of the homepage, with a workload of 3.8 user/sec;
    \item \emph{Mobile}, the request of a specific service through an API call (e.g., it maps a call from the E-Shopper mobile app), with a workload of 225 user/sec;
    \item \emph{Warehouse}, the request from a warehouse worker to set and control the availability of items, with a workload of 17.5 user/sec.
\end{itemize}

For the Train Ticket case study, we have realized two scenarios with respective workloads:
\begin{itemize}[topsep=0pt,itemsep=0pt]
    \item \emph{Rebook Ticket}, the scenario on which a customer can change a ticket reservation, with a workload of 4.5 user/sec;
    \item \emph{Update User}, the scenario on which the admin changes user information, with a workload of 2.75 user/sec.
\end{itemize}

It is worth noticing that we stimulated both the applications with workloads heavy enough to stress them, but not too heavy to originate errors and timeouts.
In order to achieve steady-state performance, for each case study we executed a 20 minutes initial warm-up with additional 10 minutes warm-up after the application of each refactoring action.
Intermediate warm-ups were necessary because the applications were never restarted to perform the refactoring actions, thus to simulate a production environment.
Response time and utilization were continuously measured for 10 minutes after (the warmup following) every refactoring action. Finally, all tests were repeated three times in order to avoid circumstantial external influence.

All the performance measurements are performed on a server with dual Intel Xeon CPU E5-2650 v3 at 2.30GHz, for a total of 40 cores and 80GB of RAM.

Performance measurements data as well as all the models resulting from this validation are available online.\footnote{Replication package: \url{https://zenodo.org/record/4756322}}

\subsection{RQ1: Performance improvement}\label{sec:evaluation:improvement}
\noindent
In order to answer RQ1, we evaluate if the approach is able to produce refactoring decisions that improve the performance metrics computed on the software models, and consequently the performance of the running system. To this end, we show how the approach applies refactoring actions that are promising on the basis of the model, and how these actions are propagated to the running system, in the context of the considered case studies. Moreover, we compare refactoring actions that were induced by performance antipatterns against other actions that we randomly perform on the model and on the running system. In this way, we are able to show that: i) our approach can select refactoring actions which improve the performance of the running system, and ii) basing the selection of such actions on the combination of QN analysis and performance antipatterns detection is more effective than randomly refactoring the system.

\begin{figure}[htbp]
    \centering
    \subfloat[Comparison of the effect of refactoring on response times.]{
        \includegraphics[height=.25\textheight]{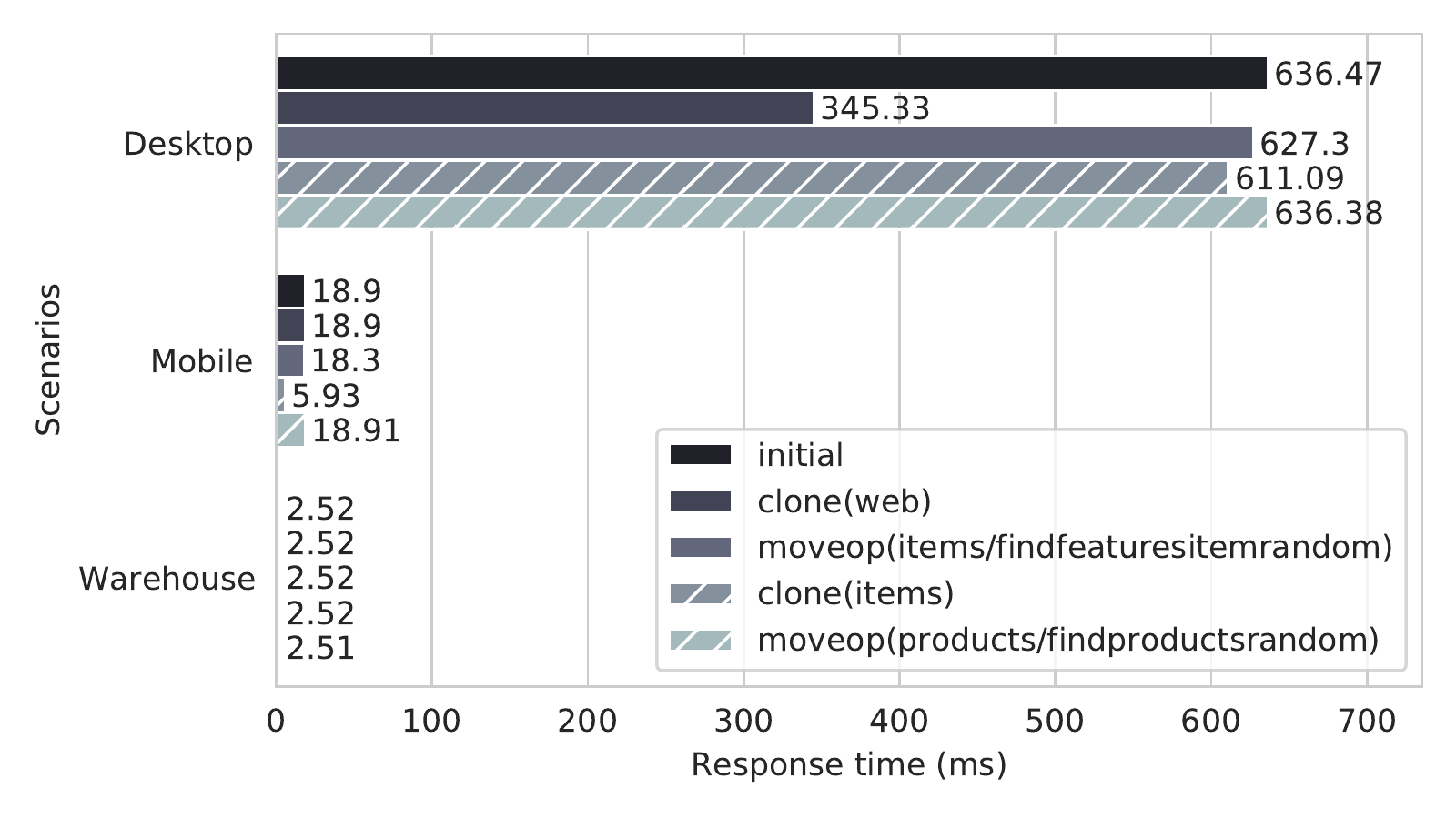}
    }
    \hspace{0.5em}
    \centering
    \subfloat[Comparison of the effect of refactoring on the utilization of microservices.\label{subfig:QN_EShopper_util}]{
        \includegraphics[height=.6\textheight]{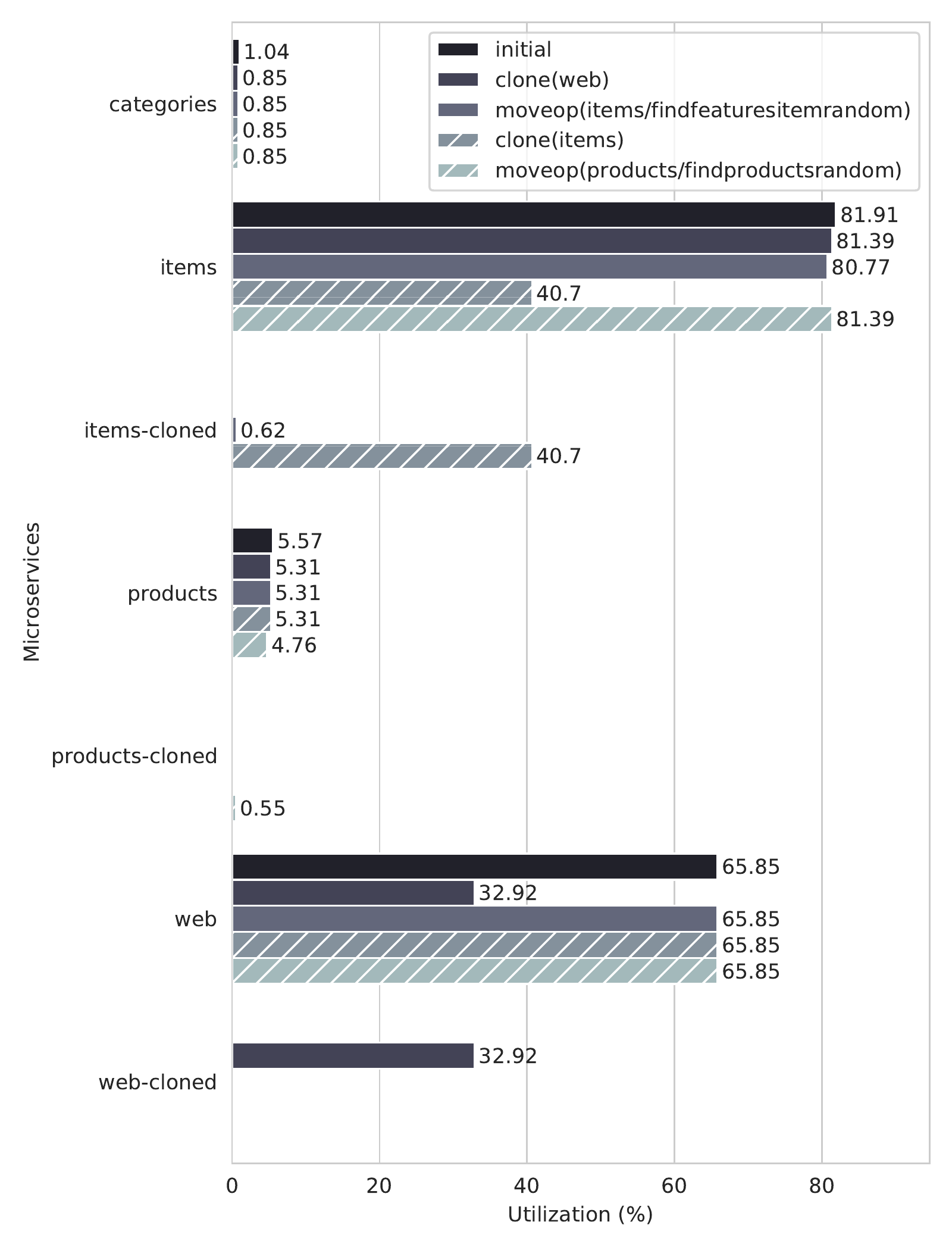}
    }
    \centering
    \caption{Average response times and utilizations computed on the QN for the E-Shopper case study.}
    \label{fig:QN_EShopper}
\end{figure}

\begin{figure}[htbp]
    \centering
    \subfloat[Comparison of the response times of scenarios.]{
        \includegraphics[height=.25\textheight]{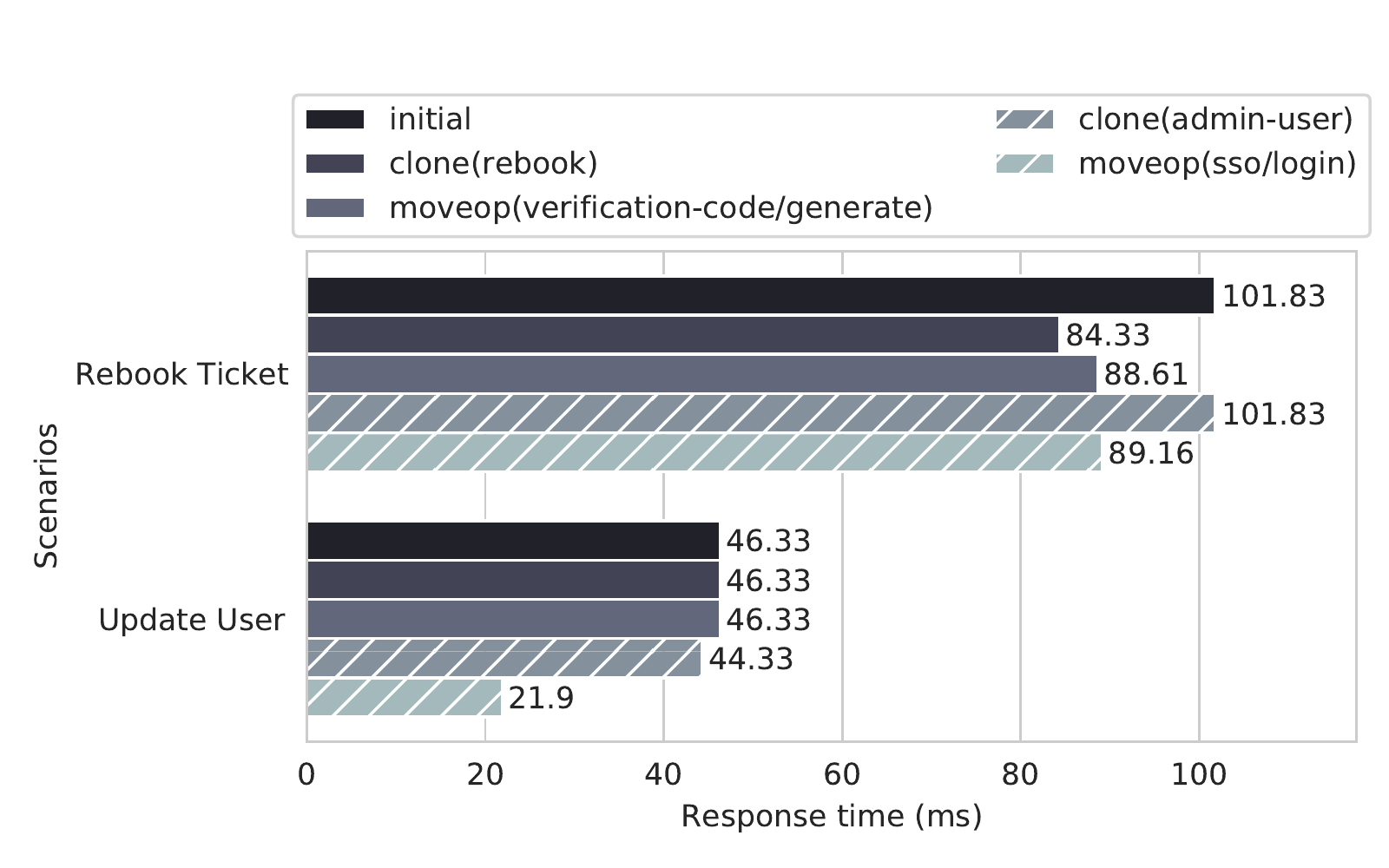}
    }
    \hspace{0.5em}
    \centering
    \begin{figrev}
    \subfloat[Comparison of the utilization of microservices.]{
        \includegraphics[height=.6\textheight]{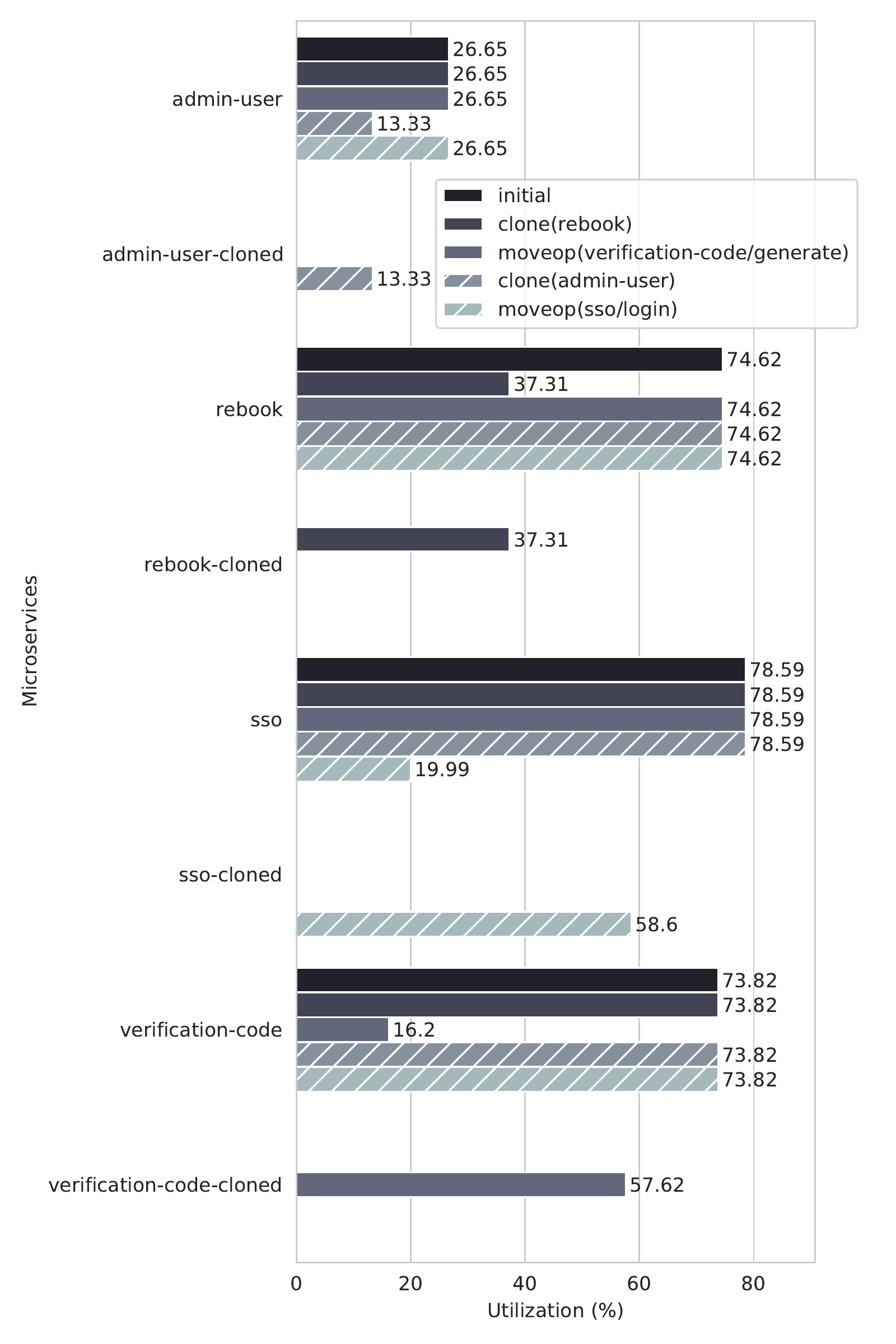}
    }
    \end{figrev}
    \caption{Average response times and utilizations computed on the QN for the Train Ticket case study.}
    \label{fig:QN_TrainTicket}
\end{figure}

We start the assessment of the refactoring impact on the model by comparing utilization and response time before and after the modifications. Figures \ref{fig:QN_EShopper} and \ref{fig:QN_TrainTicket} show a comparison of response times and utilization computed on the QN for both case studies. Utilization is reported only for relevant microservices, that are those involved in at least one scenario and for which the utilization varies among refactoring actions.

In the E-Shopper case study, starting from the initial system, our approach proposed two alternative refactoring actions to improve the \emph{Desktop} scenario: cloning the \emph{web} microservice (\emph{clone(web)}), and moving the \emph{findfeaturesitemrandom} operation from the \emph{items} microservice to a newly created one (\emph{moveop(\-items/\-findfeaturesitemrandom)}).
Among the other feasible actions, we randomly selected cloning the \emph{items} microservice (\emph{clone(items)}) and moving the \emph{findproductsrandom} operation from the \emph{products} microservice to a newly created one. These randomly selected actions are marked with a pattern in the histograms of Figure~\ref{fig:QN_EShopper}.
We can see how cloning the \emph{web} microservice is remarkably beneficial for the response time of \emph{Desktop} scenario, while the randomly selected action of the same type, that is cloning the \emph{items} microservice, has a negligible impact on the same scenario and a small impact on the \emph{Mobile} one. When comparing the actions that move an operation to a new microservice, we can notice a difference in response time, even if small, in favor of the refactoring targeting the operation \emph{findfeaturesitemrandom}.

In order to improve the \emph{Rebook Ticket} scenario of the Train Ticket case study, our approach proposed to clone the \emph{rebook} microservice (\emph{clone(rebook)}), or alternatively to move the operation \emph{generate} from the \emph{verification-code} microservice to a newly created one (\emph{moveop(verification-code/generate)}). 
The randomly selected actions are: cloning the \emph{admin-user} microservice (\emph{clone(admin-user)}), and moving the \emph{login} operation from the \emph{sso} microservice to a new one.
Also in this case, the actions selected on the basis of performance antipatterns induced a larger improvement in the response times of the targeted scenario, as we can notice from Figure~\ref{fig:QN_TrainTicket}. 

\begin{figure}[htbp]
    \centering
    \subfloat[Comparison of the effect of refactoring on response times.]{
        \includegraphics[height=.25\textheight]{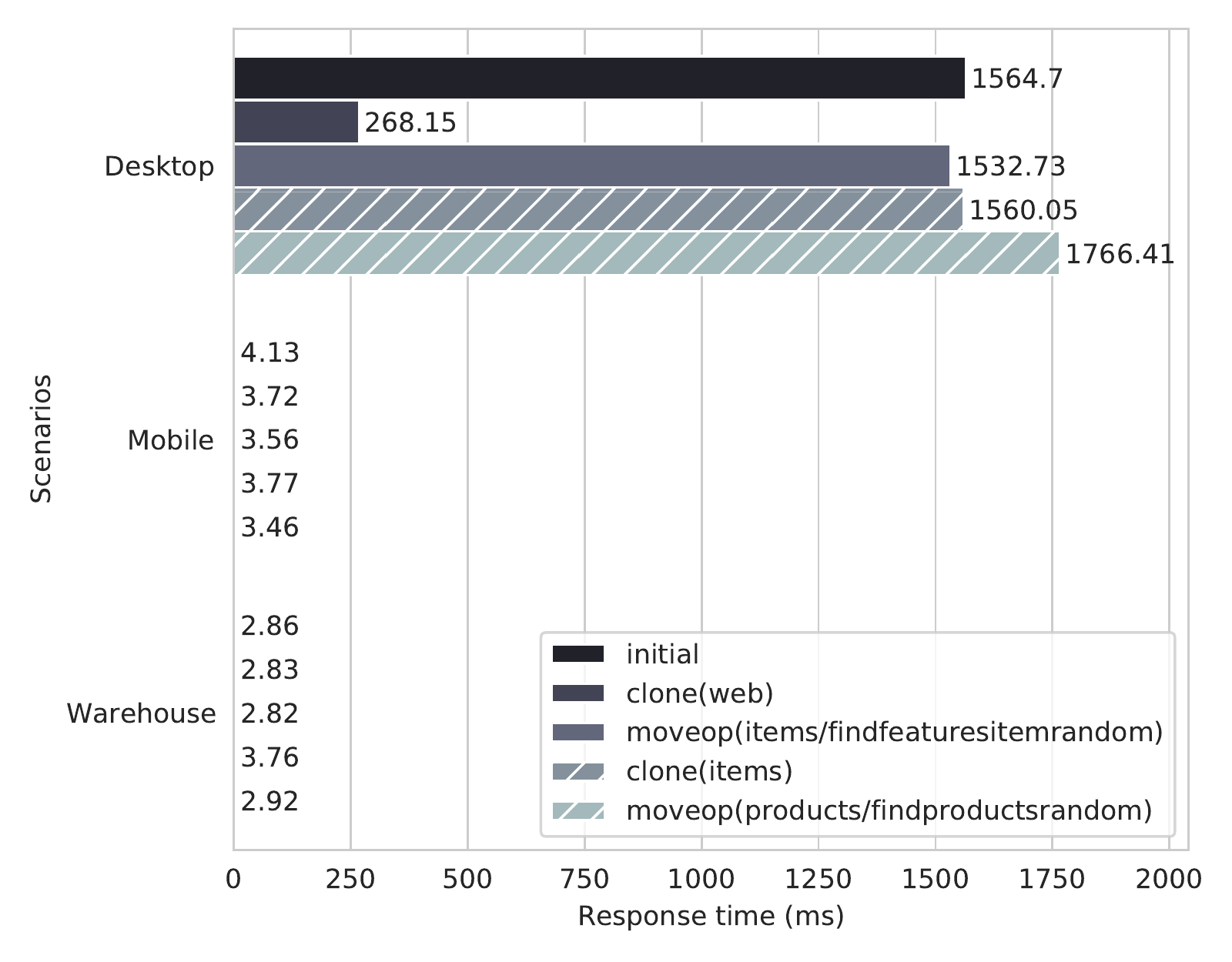}
    }
    \hspace{0.5em}
    \centering
    \subfloat[Comparison of the effect of refactoring on the utilizations of microservices.]{
        \includegraphics[height=.55\textheight]{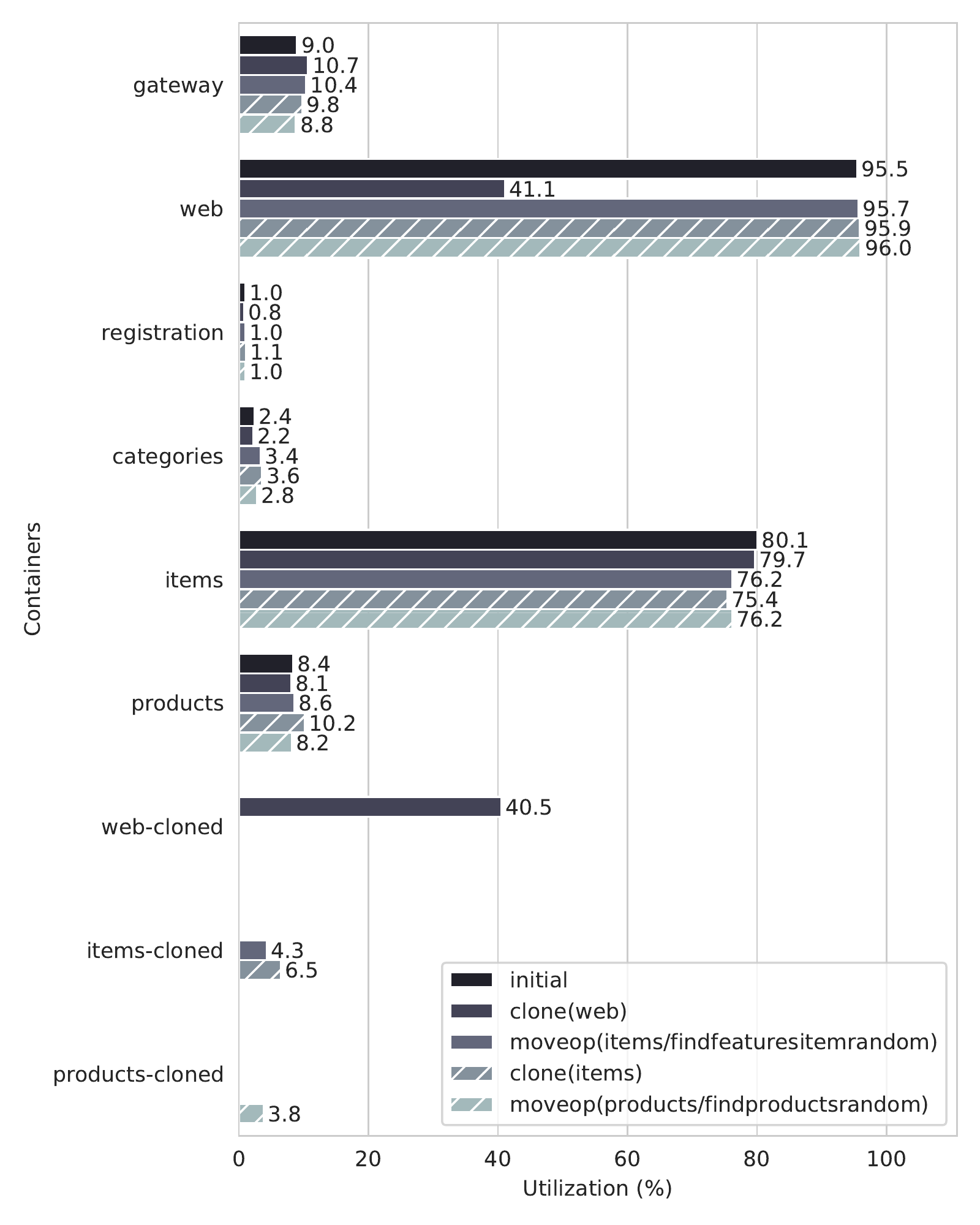}
    }
    \caption{Average response times and utilizations measured on the running system for the E-Shopper case study.}
    \label{fig:Sys_EShopper}
\end{figure}

\begin{figure}[htbp]
    \centering
    \subfloat[Comparison of the response times of scenarios.\label{subfig:Sys_TrainTicket_respt}]{
        \includegraphics[height=.25\textheight]{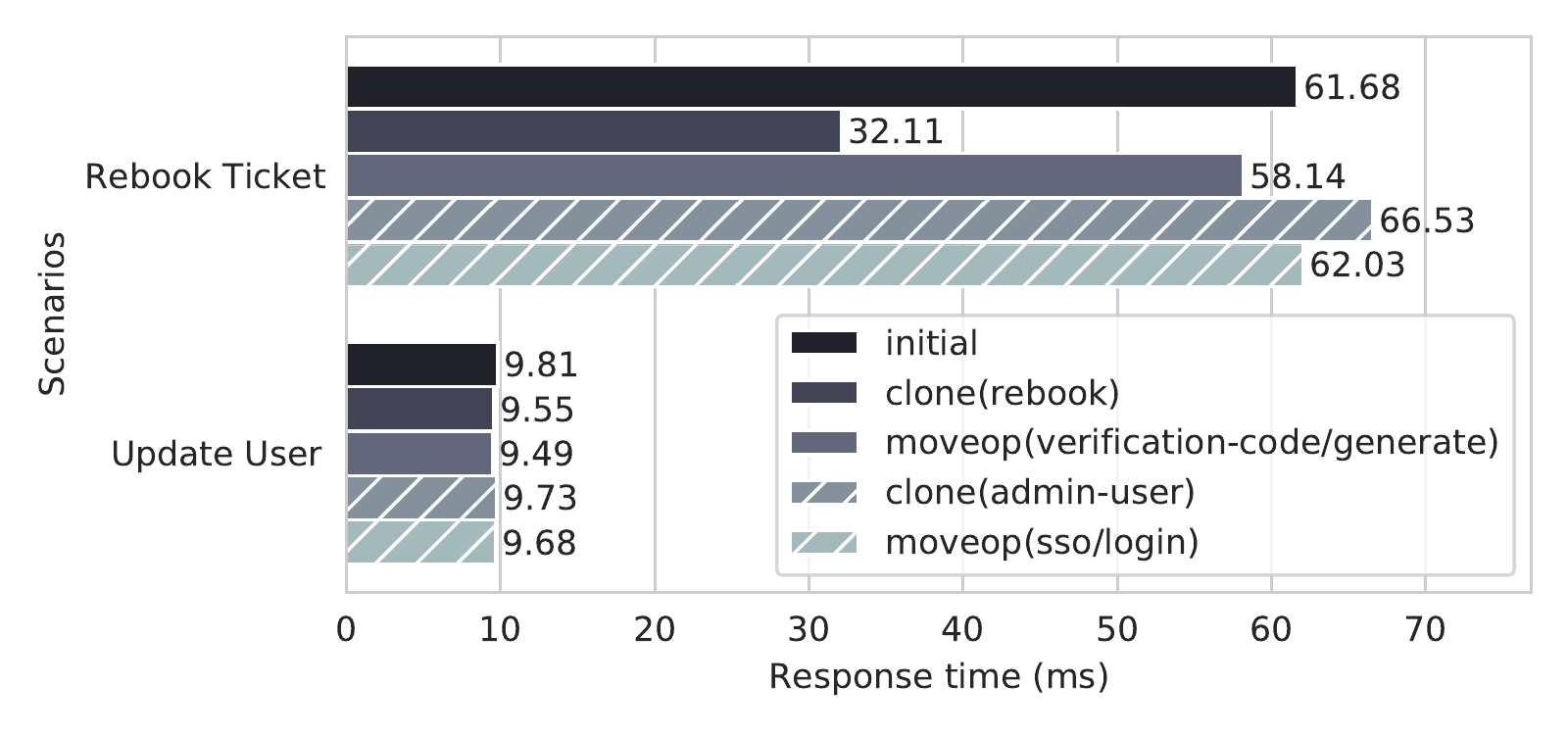}
    }
    \hspace{0.5em}
    \newline
    \centering
    \subfloat[Comparison of the utilizations of microservices.]{
        \includegraphics[height=.6\textheight]{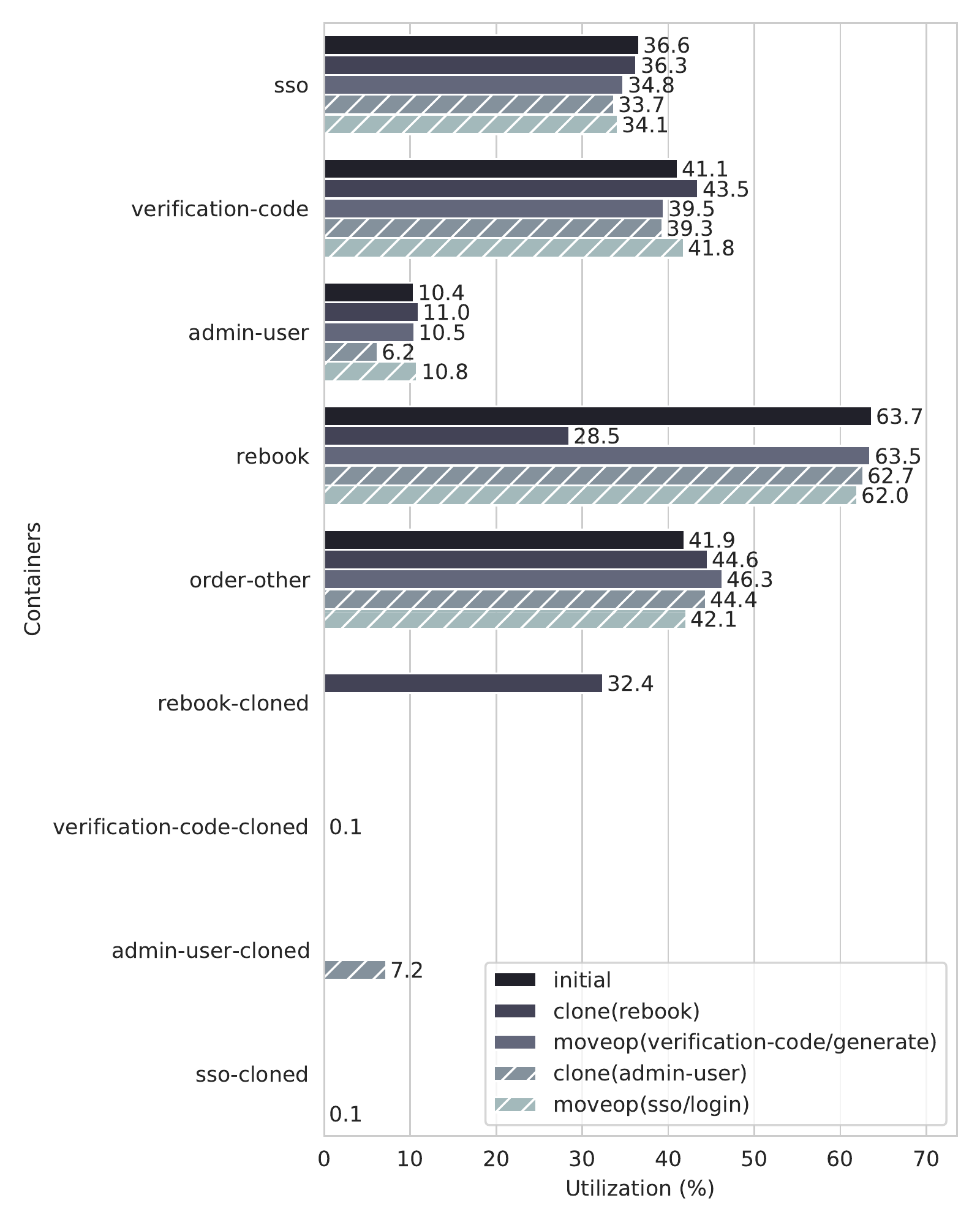}
    }
    \caption{Average response times and utilizations measured on the running system for the Train Ticket case study.}
    \label{fig:Sys_TrainTicket}
\end{figure}

As introduced before, we are also interested in evaluating if the application of the refactoring actions on the running system improves its software performance by comparing utilizations and response times before and after the modifications.
Figures \ref{fig:Sys_EShopper} and \ref{fig:Sys_TrainTicket} show the measures obtained by monitoring both case studies as described in Section~\ref{subsec:monitoring_techniques}.

The results show that, in both case studies, the refactoring actions that were selected on the basis of antipatterns are more effective in improving the response times of the targeted scenarios.
Furthermore, there are some interesting aspects to notice.
For instance, by just looking at the utilizations computed on the QN for the E-Shopper case study (Figure~\ref{subfig:QN_EShopper_util}), a performance analyst would have probably guessed that cloning the \emph{items} microservice would be the best action to perform. Instead, such refactoring only marginally decreases the response time of the \emph{Desktop} scenario on the QN and on the system.
In this case, cloning the \emph{web} microservices was far more effective as also shown by the measures obtained from the running system. 

We can notice an even more extreme situation in the Train Ticket case. Both the actions that were randomly selected actually increased the response time of the \emph{Rebook Ticket} scenario on the running system (Figure~\ref{subfig:Sys_TrainTicket_respt}).
Even if, in general, the cloning and moving operation refactoring actions are designed to produce a performance improvement, when applied without taking into consideration the design of the application may indeed result in degrading the performance.

These are just some examples of how the combination of design and runtime knowledge can induce a more thorough selection of the convenient refactoring to perform. More generally, design-runtime traceability provides additional knowledge that can be automatically maintained while supporting design decisions, also when the software is running in production.

\subsection{RQ2: Performance Antipatterns}\label{sec:evaluation:pas}

\noindent
In order to answer RQ2, we discuss the benefits of employing detection and refactoring of PAs to performance improvement forecasting. 
We consider here performance antipatterns that can suitably fit with micro\-service-based systems, namely Blob and PaF (Pipe and Filter)~\cite{Smith:2002vl}.

As described in Section~\ref{sec:background:pas}, literals of a performance antipattern first-order logic representation are compared to thresholds. The definition of fixed values for these thresholds is an application-dependent task that can become very complex in some cases. Therefore, we employ in this paper the concept of fuzzy thresholds~\cite{DBLP:conf/fase/ArcelliCT15}.  Instead of deterministically identifying the occurrence of a performance antipattern, fuzziness in thresholds induces a probability for an antipattern to occur, as the combination of probabilities of threshold violations. The probability for an element $x$ to violate a fuzzy threshold $Th_k$ on $k$ metric is defined as follows: 
\begin{equation}
    P_{k}(x) = 1 - \frac{UBTh_{k} - F_{k}(x)}{UBTh_{k} - LBTh_{k}}
    \label{eq:fuzzy-prob}
\end{equation}

This expression considers $UPTh_k$ and $LBTh_k$ as the upper and lower bounds of $Th_k$ threshold, respectively, and $F_{k}(x)$ as the value that the $k$ metric assumes in the $x$ element. 

\paragraph{Blob}

Each inequality of Expression~\ref{eq:pas.blob}, described in Section~\ref{sec:background:pas}, undergoes a fuzzy evaluation like the one in Expression~\ref{eq:fuzzy-prob}. Hence, the probability of a Blob occurrence is obtained as follows:
\begin{equation}
P(Blob) = P_{numClientConnects}(C_x) * P_{numMsgs}(C_x, C_y) * P_{maxHwUtil}(P_{xy})
\label{eq:fuzzy-blob}
\end{equation}

\paragraph{Pipe and Filter (PaF)}

Each inequality of Expression~\ref{eq:pas.paf}, described in Section~\ref{sec:background:pas}, also undergoes a fuzzy evaluation like the one in Expression~\ref{eq:fuzzy-prob}. Hence, the probability of a PaF occurrence is obtained as follows:
\begin{equation}
P(PaF) = P_{resDemand}(Op) * P_{maxHwUtil}(P) 
\label{eq:fuzzy-paf}
\end{equation}

We remark that the second literal in Expression~\ref{eq:pas.paf} must be always equal to one to trigger a PaF, thus it can be omitted in Expression~\ref{eq:fuzzy-paf} that quantifies the PaF occurrence probability.
\qed

In the following, \tabref{tab:evaluation.eshopper.blob.web}~--~\tabref{tab:evaluation.tt.blob.verification} describe the effects on performance antipattern probabilities of refactoring actions that are either randomly selected or driven by performance antipattern detection. In each table, the $M_0$ column lists the initial probability, \ie the value obtained on the initial configuration, the $M_1$ column lists the probability obtained after applying the action suggested by PADRE to remove the antipattern, and the $M_2$ column lists the probability obtained after applying a random refactoring action. Each row represents a literal of the performance antipattern expression, while the last row reports the occurrence probability of the whole antipattern.

\subsubsection{E-Shopper}

 \tabref{tab:evaluation.eshopper.blob.web} and \tabref{tab:evaluation.eshopper.blob.items} describe the probability of \emph{Web} and \emph{Items} being Blob performance antipatterns, respectively, in the E-Shopper case study. In particular, column $M_1$ of \tabref{tab:evaluation.eshopper.blob.web} corresponds to the \emph{Clone} refactoring action on \emph{Web} microservice as suggested by PADRE, while column $M_2$ corresponds to the \emph{Clone} refactoring action on \emph{Items} as a randomly selected recfactoring action. 
It is noteworthy that the probability drops from $0.80$ to $0.39$ after the refactoring action suggested by PADRE, while the probability grows to $1$ after the random action of \emph{Items} cloning. 

We recall that the application of the same refactoring actions on the running system, as shown in \secref{sec:evaluation:improvement}, in case of cloning \emph{Web} reduces the utilization of that microservices by 50.2\%, whereas cloning \emph{Items} does not change the utilization of the Web microservice.

Although the initial probability of \emph{Items} being a \emph{Blob} is lower than in the \emph{Web} case, the \emph{Move Operation} on \emph{items/findfeaturesitemrandom}, suggested by PADRE, reduces the initial probability ($M_0$) from $0.25$ to $0.06$, while the application of the \emph{Move Operation} on the randomly selected \emph{products/findproductsrandom} does not change the probability at all. 

We recall that in the running system, as shown in \secref{sec:evaluation:improvement}, the effect of the PADRE suggestion is negligible in terms of performance, as the utilization of \emph{Items} is reduced by 0.4\% and the response time (see \figref{fig:Sys_EShopper}) is quite the same. However, the effect of the random action in terms of performance is twofold: the utilization of \emph{Items} is decreased by 4\%, and the utilization of \emph{Web} is increased, albeit it is close to being saturated.

\begin{table}[ht]
\centering
 \scriptsize
    \begin{tabular}{llccc}
    \toprule
    \multicolumn{2}{c}{Performance Antipattern Literal } & $M_0$   & $M_1$ & $M_2$  \\
    \midrule
    \multicolumn{2}{l}{$P_{numClientConnects}(Web)$}            &   1      & 1     & 1 \\ 
    \multicolumn{2}{l}{$P_{numMsgs}(Web)$}                      &   1      & 1     & 1 \\ 
    \multicolumn{2}{l}{$P_{maxHwUtil}(Container-Web)$}          &   .80    & .39   & 1 \\
    \midrule
    \multicolumn{2}{l}{$P_{Blob}(Web)$}             &   .80    & .39   & 1 \\
    \bottomrule
    \end{tabular}
    \caption{\label{tab:evaluation.eshopper.blob.web}Probability of \emph{Web} being a Blob. $M_0$ is the initial model, $M_1$ is the model refactored through a \emph{Clone} refactoring action on \emph{Web}, and $M_2$ is the model refactored through a \emph{Clone} refactoring action on \emph{Items}.}
\end{table}

\begin{table}[ht]
\centering
 \scriptsize
    \begin{tabular}{llccc}
    \toprule
    \multicolumn{2}{c}{Performance Antipattern Literal } & $M_0$   & $M_1$  & $M_2$  \\
    \midrule
    \multicolumn{2}{l}{$P_{numClientConnects}(Items)$}          &   .5     & .25   & .5  \\ 
    \multicolumn{2}{l}{$P_{numMsgs}(Items)$}                    &   .5     & .25   & .5  \\ 
    \multicolumn{2}{l}{$P_{maxHwUtil}(Container-Items)$}        &   1      & 1     & 1   \\
    \midrule
    \multicolumn{2}{l}{$P_{Blob}(Items)$}             &   .25    & .06   & .25 \\
    \bottomrule
    \end{tabular}
    \caption{\label{tab:evaluation.eshopper.blob.items}Probability of \emph{Items} being a Blob. $M_0$ is the initial model, $M_1$ is the model refactored through a \emph{Move Operation} refactoring action on \emph{items/findfeaturesitemrandom}, and $M_2$ is the model refactored through a \emph{Move Operation} refactoring action on \emph{products/findproductsrandom}.}
\end{table}

\subsubsection{Train Ticket}

We report in \tabref{tab:evaluation.tt.paf.verification}~--~\tabref{tab:evaluation.tt.blob.verification} performance antipattern probabilities due to different refactoring actions applied to the Train Ticket case study.

\tabref{tab:evaluation.tt.paf.verification} reports the probability of \emph{verification-code/generate} operation being a Pipe and Filter (PaF). In particular, column $M_1$ refers to a \emph{Move operation} refactoring action on \emph{verification-code/generate} suggested by PADRE, while column $M_2$ refers to a randomly chosen \emph{Move Operation} refactoring action effects on \emph{sso/login} operation. We notice that the PaF probability decreases from $0.80$ to $0$ by applying the refactoring action suggested by PADRE. If we look at the running code (see \figref{fig:Sys_TrainTicket}), this refactoring action decreases the utilization of \emph{Container-Verification} microservice as well as the response time of the \emph{Rebook Ticket} scenario.
The random action, instead, increases the probability of \emph{verification-code/generate} being a PaF to $0.87$, and if we look at the running system, this refactoring action leads to increase both the utilization of \emph{Container-Verification} and the response time of \emph{Rebook Ticket} scenario.

\begin{table}[ht]
\centering
 \scriptsize
    \begin{tabular}{llccc}
    \toprule
    \multicolumn{2}{c}{Performance Antipattern Literal }      & $M_0$ & $M_1$ & $M_2$ \\
    \midrule
	\multicolumn{2}{l}{$P_{resDemand}(verification-code/generate)$} & .88   & .88   & .88   \\
	\multicolumn{2}{l}{$P_{maxHwUtil}(Container-Verification)$}     & .91   & 0     & .98   \\
	\midrule 
	\multicolumn{2}{l}{$P_{PaF}(verification-code/generate)$}                  & .80   & 0     & .87   \\
    \bottomrule
    \end{tabular} 
    \caption{\label{tab:evaluation.tt.paf.verification}Probability of \emph{verification-code/generate} being a Pipe and Filter (PaF). $M_0$ is the initial model, $M_1$ is the model refactored through a \emph{Move operation} refactoring action on \emph{verification-code/generate}, and $M_2$ is the model refactored through a \emph{Move operation} refactoring action on \emph{sso/login}.}
\end{table}

\tabref{tab:evaluation.tt.blob.rebook} reports the probability of \emph{rebook} being a Blob. In particular, column $M_1$ lists the probabilities related to the \emph{Clone} refactoring action on \emph{rebook}, as PADRE suggests, while the column $M_2$ lists the probabilities related to the same refactoring action on \emph{admin-user}. We notice that probability decreases from $0.46$ to $0.12$ by applying the refactoring action suggested by PADRE, while it remains unchanged after the randomly chosen action. If we look at the running code (see \figref{fig:Sys_TrainTicket}), the effect of the random action of cloning the \emph{admin-user} microservice on the utilization is negligible, but it increases the response time of the \emph{Rebook Ticket} scenario by 7\%. Instead, the effect of the refactoring action suggested by PADRE on system performance is twofold: i) the response time of the \emph{Rebook Ticket} scenario decreases by 47.9\%, and ii) the utilization of the \emph{Container-Rebook} microservice decreases by 55\%.

\begin{table}[ht]
\centering
 \scriptsize
    \begin{tabular}{llccc}
    \toprule
    \multicolumn{2}{c}{Performance Antipattern Literal }   & $M_0$ & $M_1$ & $M_2$ \\
    \midrule
    \multicolumn{2}{l}{$P_{numClientConnects}(rebook)$}          &   .5  & .5    & .5  \\ 
    \multicolumn{2}{l}{$P_{numMsgs}(rebook)$}                    &   1   &  1    & 1   \\ 
    \multicolumn{2}{l}{$P_{maxHwUtil}(Container-Rebook)$}        &  .92  & .24   & .93 \\
	\midrule
	\multicolumn{2}{l}{$P_{Blob}(rebook)$}               &  .46  & 0.12  & .46 \\
    \bottomrule
    \end{tabular} 
    \caption{\label{tab:evaluation.tt.blob.rebook}Probability of \emph{rebook} being a Blob. $M_0$ is the initial model, $M_1$ is the model refactored through a \emph{Clone} refactoring action on \emph{rebook}, and $M_2$ is the model refactored through a \emph{Clone} refactoring action on \emph{admin-user}.}
\end{table}

\tabref{tab:evaluation.tt.blob.verification} reports the probability of \emph{verification} being a Blob. In particular, column $M_1$ lists the probabilities of the \emph{Move Operation} refactoring action on \emph{verification-code/generate}, as PADRE suggests, while column $M_2$ reports the probabilities of the same refactoring action on \emph{sso/login}. We notice that the probability decreases from $0.46$ to $0$ by applying the refactoring action suggested by PADRE, which also reduces the \emph{Verification.code} utilization and the response time of the \emph{Rebook Ticket} scenario (as shown in Figure \ref{fig:Sys_TrainTicket}).
The application of the random action instead increases the probability of \emph{verification-code} being the Blob to $0.98$.

\begin{table}[ht]
\centering
 \scriptsize
    \begin{tabular}{llccc}
    \toprule
    \multicolumn{2}{c}{Performance Antipattern Literal}   & $M_0$ & $M_1$ & $M_2$ \\
    \midrule
    \multicolumn{2}{l}{$P_{numClientConnects}(verification-code)$}    &   .5  &  0    & 1  \\ 
    \multicolumn{2}{l}{$P_{numMsgs}(verification-code)$}              &   1   &  0    & 1   \\ 
    \multicolumn{2}{l}{$P_{maxHwUtil}(Container-Verification)$}  &  .91  &  0    & .98 \\
	\midrule
	\multicolumn{2}{l}{$P_{Blob}(verification-code)$}              &  .45  &  0    & .98 \\
    \bottomrule
    \end{tabular} 
    \caption{\label{tab:evaluation.tt.blob.verification}Probability of \emph{verification-code} being a Blob. $M_0$ is the initial model, $M_1$ is the model refactored through a \emph{Move Operation} refactoring action on \emph{verification-code/generate}, and $M_2$ is the model refactored through a \emph{Move Operation} refactoring action on \emph{sso/login}.}
\end{table}

\subsection{Summarizing discussion}
On the basis of the results obtained, we can state that
the removal of performance antipatterns leads to a performance improvement of a running system. On the other hand, ignoring the performance antipatterns knowledge induces either unchanged performance, in the best case, or performance detriment in all other cases.

Indeed, we experience on the Train Ticket case study a  performance improvement in terms of lower hardware utilization from 0.411 to 0.395 (\ie by about 4\%)
when the \emph{verification-code/generate} operation is moved, and the response time of the \emph{Rebook Ticket} scenario is reduced by 5\% as well. It is noteworthy that the same refactoring action also removes the ``Pipe and Filter'' performance antipattern (as shown in \tabref{tab:evaluation.tt.paf.verification}).

Also, we experience on the E-Shopper case study a lower hardware utilization from 0.955 to 0.411 (\ie by about 57\%) when the \emph{Container-Web} is cloned (as suggested by PADRE), and the response time of the \emph{Desktop} scenario is reduced by about 83\% (\ie from 1,564 ms to 268 ms as shown in \figref{fig:Sys_EShopper}), and the probability of Web being a ``Blob'' performance antipattern decreases by 39\%, as shown in \tabref{tab:evaluation.eshopper.blob.web}.

We have also noticed that the application of a random refactoring action induces either unchanged performance or performance detriment. For example, the Train Ticket case study has shown a higher utilization from 0.411 to 0.418 (\ie an increment by 1.7\%), while the response time of the \emph{Rebook Ticket} scenario remained unchanged. The probability of \emph{verification-code/generate} being a ``Pipe and Filter'' performance antipattern is increased by 11\% (as shown in \tabref{tab:evaluation.tt.paf.verification}. 
Instead, the random action on the E-Shopper case study has caused no changes both for the utilization, the response time, and the probability of \emph{Web} being a ``Blob'' performance antipattern (as shown in \tabref{tab:evaluation.eshopper.blob.web}).

  \section{Threats to validity}\label{sec:t2v}

In this section, potential threats to validity associated with the validation are discussed.

\emph{Conclusion validity} concerns the reliability of the measures. As explained in Sec. \ref{sec:evaluation:setup}, we properly and rigorously designed our case studies setup. We attempted to avoid any bias by: i) generating different scenarios with respective workloads to simulate different parts of the systems; ii) stressing the execution of the applications to check performance under realistic conditions; iii) repeating the measurements several times in order to avoid circumstantial external influences.
We cannot assert how realistic the selected scenarios are, because these applications have been mostly used in literature for sake of validation in controlled environments. For example, in \cite{Kounev21} Train Ticket has been used by exposing numerous services to a base load that varies between 3 and 22 requests per second, and the workload that we have adopted in this paper for the same application falls within this range. However, all the adopted application settings are among the ones that lead the considered applications on performance boundary situations that enable to validate the effectiveness of the approach proposed here.

Furthermore, during the evaluation, we ensured that the observed performance improvement was actually induced by the refactoring actions selected by our approach. To this end, we compared our solutions with performance variations caused by randomly selected refactoring actions.

\emph{Internal validity} concerns any extraneous factor that could influence our results. In general, the implementation of the approach could be defective, as well as the results of the analysis could be inaccurate. 
In order to avoid any bias: i) we have assured that the implementation was aligned with the software design to generate accurate traceability models and consistently annotate the performance models; ii) we completely delegated the analysis of the considered performance model to an external consolidated tool; iii) we employed technologies that are widely tested in production (e.g., Zuul, Eureka, Nginx) that provide unified interfaces to allow refactoring of microservices without exposing the internal structure.
Furthermore, we provided a detailed discussion of the code instrumentation and made publicly available the source code in order to allow other researchers to reproduce and inspect our results.

\emph{Construct validity} concerns any factor that can compromise the validity of the measurements and the resulting observations. 
Evaluation results are highly dependent on the quality of the considered measures. In the evaluation, we apply monitored data observed from running system and performance measures.
As described above, we properly designed our case studies setup to avoid influence factors. For instance, we disabled as many system services (Linux \emph{systemd} units) as possible to lessen the effects of context switching. To avoid interferences that may be caused by internal errors, we also monitored the machine by looking at system logs (\emph{dmesg} buffer and \emph{systemd} journal) for unexpected entries.Another threat may be posed by missing links in the generated traceability models. This may be caused by errors in the generation of Log models from monitoring traces, or by incorrectly matching the patterns defined in the correspondences specification, either for UML or Log models. To avoid this, we performed several manual assessments of the generated traceability models by looking for missing links or links connecting the wrong elements.
Finally, we have annotated the performance indices in the UML-MARTE models in a consistent way with the consolidated literature in the field of software performance.

\emph{External validity} refers to the generalizability of the obtained results.
Case studies may be selected to facilitate a deeper understanding of the approach and this could affect their representativeness. In order to mitigate this, we selected two microservice-based web application benchmarks that have been successfully used in previous work for their size and heterogeneity~\cite{ACPET19,DBLP:conf/staf/Pompeo0CE19}. 
This choice also provides indicators for cases having similar properties. 
With reference to the used languages and tools, we adopted specific development and monitoring technologies (i.e., based on the Spring framework), as well as the specific modeling standards like UML and MARTE. These technologies are widely used both in academia and industry, thus supporting the approach to be generalizable and easily reproduced in other contexts.   

Adopting specific tools (JTL and PADRE) could threat the generalizability of the approach. In this merit, very few other alternative tools are available for replacing JTL, and actually none for replacing PADRE to detect performance antipatterns in UML models. However, both tools have proven to be suitable for the specific purposes of the approach and to be used in heterogeneous contexts.

 \section{Related Work}\label{sec:related}

\noindent
In this section we discuss existing work that proposes approaches for archi\-tectural-based improvement of software systems driven by the observation of monitoring data and/or their interleaving with the software design modeling. Researchers from several areas (\eg self-adaptive systems, software engineering and continuous system engineering) have actively studied a wide variety of methods and techniques applicable at design time and/or at runtime. Hereafter, we focus on approaches dedicated to software performance and on approaches that make use of software models in the domain of microservices.  

\subsection{Software performance engineering approaches}

\noindent
A vast literature exists on performance modeling, performance monitoring, and performance problem identification techniques, as quite separate research domains. We report on the most significant papers that attempt at merging such domains.

Trubiani et al. \cite{DBLP:journals/infsof/TrubianiBHAK18} have provided a systematic process to identify performance issues from runtime data, based on load testing coming from operational profile and application profiling. In particular, from runtime data, performance antipatterns are detected, aimed at identifying common performance issues and their solutions. Software refactoring is then (manually) applied to solve identified performance antipatterns. 
Apart from technological and implementation aspects, a methodological difference distinguishes our approach from the one in \cite{DBLP:journals/infsof/TrubianiBHAK18}, namely: we bring runtime data up to the design level, by annotating a UML model with the MARTE profile, and one of the main advantage of addressing performance issues at design level, among other, is to narrow down the search space for potential actions that can beneficially affect system performance. 

Menasc{\'{e}} et al. \cite{DBLP:conf/models/PorterMG16} have proposed the DeSARM approach, whose scope is the derivation of architectural models at runtime. Such models can be used in decentralized decision-making for architecture-based adaptation in large distributed systems. To this aim, DeSARM is able to identify important architectural characteristics of a running application, such as components, connectors, nodes and  communication patterns. 
DeSARM has been used by Albassam et al. \cite{DBLP:conf/models/GomaaA17} to introduce runtime failure analysis and architectural recovery on the discovered system architecture. However, DeSARM has not been adopted for identifying performance problems, as we do in this paper. 

Petriu et al. \cite{Altamimi:2016:PAR:3049877.3049899} deal with the automated generation of performance models from UML-MARTE architectural models, and the propagation of performance analysis results back to the latter. 
The main difference with our work is that this approach fully works at the model level, and it does not consider the availability of a running software system from which runtime information can be extracted for a more accurate identification of performance problems.

Logs obtained from monitoring a running software system are exploited by van Hoorn et al. \cite{Vogele2018} to automatically extract workload specifications for load testing and performance models parameterization. However, \cite{Vogele2018} is limited to the parameterization of performance models, whilst our approach provides support for the interpretation of analysis results carried out by performance antipattern detection and their possible solutions. 

Mazkatli et al. \cite{Mazkatli20} have presented an approach for continuous integration of performance models that considers parametric dependencies after analyzing the source code changes. The approach builds upon the Palladio approach and the goal is to automatically keep the performance models up-to-date to allow architecture-based performance prediction. In contrast with this approach, we have provided an approach to identify model-based design alternatives to overcome detected performance problems and to apply them on the system. 

Recently, Heinrich \cite{Heinrich20} has proposed an approach to align architectural models used in development and operation by means of a correspondence model between implementation artifacts and component-based Palladio architectural models. 
In contrast with this work, our approach uses UML that is a widely adopted standard respect to the Palladio Component Model. They use an ad-hoc correspondence model that is then exploited by a complex pipeline of transformations, while we propose a general approach where the correspondences are generated from a declarative specification easily adaptable to other contexts. They use a specific infrastructure monitoring, while we combine application instrumentation and infrastructure monitoring. Also the scope is different, as their contribution is to assess software performance with the aim to support design decisions, whereas our approach is not limited to build design-runtime correspondences and predict performance issues. We indeed enable the identification of design alternatives and the implementation of system refactoring actions to improve performance. 
On the other hand, in contrast with our approach, workload characterization and model structure updates are proposed in \cite{Heinrich20}, where the model is updated online and scalability and accuracy are validated. Finally, it is not limited to microservices domain like our approach does.

\subsection{Model-based approaches for microservices}

\noindent
In self-adaptive systems, software models have been mostly used at development and specification time, and a few works considered software models for microservice application adaptations. In this respect, Rademacher \cite{RademacherSZ17} surveyed the use of models in microservice and service-oriented architectures. Also, Derakhshanmanesh \cite{Derakhshanmanesh16} provided a vision and future challenges on the use of domain-specific modeling languages and model transformations across the full software lifecycle (including runtime) to define and evolve a microservice application at the architectural level.

Weyns \cite{DBLP:books/sp/19/Weyns19} described relevant aspects and future challenges in the field of software engineering for self-adaptive systems. In particular, the author puts the concrete realization of runtime adaptation mechanisms that leverage software models at runtime to reason about the system and its goals. The use of models at runtime (known as models@run.time)~\cite{modelsruntime,Bencomo2019} has been proposed to extend the applicability of software models produced in MDE approaches to the runtime environment. Such models should represent the system and its current/updated state and behavior. The envisioned goal is to support adaptive systems, e.g., to drive subsequent adaptation decisions, to fix design errors or to explore new design decisions. As an alternative to models at runtime, we used traceability models to represent runtime information and its relation with design models. Such solutions allow us to exploit existing monitoring infrastructure and existing design models throughout all phases of the approach. 

Dullmann and van Hoorn \cite{DullmannH17} proposed a preliminary framework to generate microservice environments that can then be used for measurement-based evaluation of performance and resilience. The approach allows developers to create models and generate Java code and deployment files. The generated microservices are automatically instrumented to collect metrics at runtime. In contrast with our work, the proposed setup was developed and used  as benchmarking environments for their evaluation of approaches for performance and resilience. Although the framework is able to generate microservice environments with specified properties, it has not been adopted for the adaptation of microservice-based systems, as we do in this paper. 

Zuniga-Prieto et al. \cite{Prieto2017}  proposed an incremental and model-driven approach that supports the integration of cloud service applications and their dynamic architecture reconfiguration.
Models are used to generate skeletons of microservices, integration logic, and also scripts to automatically deploy and integrate the microservices in the specific cloud environment where a microservice will be deployed. The approach supports the integration of increments composed of several microservices, whereas it has not been adopted for the improvement of existing services; also, runtime information is not considered.  

Sampaio et al. \cite{Sampaio2019} proposed  a platform-independent runtime adaptation mechanism to reconfigure the placement of microservices based on their communication affinities and resources usage. The authors propose to identify the runtime aspects of microservice execution that impact the placement of microservices by using models at runtime.  In contrast with our approach, the main contribution is limited to the reconfiguring mechanism to manage the placement of microservices.

 \section{Conclusion}\label{sec:conclusion}

\noindent
In this paper we have extended a previously introduced approach \cite{ACPET19} aimed at supporting the identification and solution of performance problems on a running system by 
means of an integrated approach that exploits traceability relationships between the monitoring data and the architectural model to detect and solve performance antipatterns.
The extensions introduced in this paper have consisted of: 
i) collecting a larger set of metrics and performance measures from the running systems;
ii) translating refactoring actions (suggested by performance model analysis) into refactorings applied to a running system;
iii) closing the performance engineering loop on the original E-Shopper case study by working on its running system;
iv) fully applying the whole approach to the additional Train Ticket case study (at the model and running system level).

The application of our approach to two case studies has consolidated the insight that the definition and the use of traceability models is a key point for automating the exploitation of runtime information for sake of performance analysis and improvement. In fact, in absence of a collection of correspondences between monitored data and system design, the extrapolation of a large amount of data from a running application and the annotation of a software model would be very expensive.

By validating our approach also on Train Ticket, which is a benchmark of realistic size, we have realized that the application of refactoring actions at runtime is not only possible, but also easy to adjust on different technologies.
The merit for this result may probably be found in the key concepts imposed by the microservices paradigm, like having small and focused components and emphasizing composability.

The extension of our approach to a larger set of performance metrics confirmed that design/runtime interactions should be automatically generated from a clear declarative specification.
In this way, the introduction of new monitoring measures, or even different modelling notations, is just a matter of changing the name and the format of the elements to be matched by the correspondences specification.
And this is evidently more convenient than relying on general purpose code to generate such correspondences.

Only by applying refactoring actions on a running system we have been able to appreciate the crucial role of design knowledge to the selection of those actions that better improve a target performance metric. Such knowledge may be proven very difficult to reconstruct just from monitoring. Indeed, while directly extracting performance models from runtime information may obtain accurate models in terms of the ability to predict a performance change, it could be limiting. In fact, by discarding design models, such approaches are usually compelled to reversely engineered information, and they do not look at already available information, like the role of a component in a scenario or how intensely an operation is supposed to be invoked.

An interesting direction for future work would be to compare the results obtained by applying the refactoring actions suggested by our approach with the ones that would have been obtained by applying human suggestions in the same situations. Indeed, a fully automated process is difficult to be accepted in actual industrial contexts, where the human expertise could be exploited, for example, when multiple refactoring options occur.

We conceived and implemented the approach to be extensible. In this paper, we defined a dedicated metamodel that represents the logs format as represented by the used monitoring infrastructure. However, runtime information can be of different types (e.g., simulation or executable models, logs/traces, states or configurations of the system, test models, dynamic information or runtime measures), it can have different formats (e.g., textual, binary, datasets) and can be collected by means of various mechanisms (e.g., simulation, monitoring, execution, debugging, profiling, verification). As an enhancement, a generic log metamodel that deals with different runtime information can be specified and integrated to the approach. 

As a further extension, also the adoption of different modeling languages can be supported by the approach and by the used tools. The designer can specify the correspondences between proper languages; JTL is indeed able to deal with any Ecore artifact conforms to EMF. Moreover, we envisage possibly redefining the model annotation step and the antipattern detection rules. PADRE indeed provides the designer with interfaces to write proper notation-specific rules.

\bibliography{biblio}

\begin{thebibliography}{10}
\expandafter\ifx\csname url\endcsname\relax
  \def\url#1{\texttt{#1}}\fi
\expandafter\ifx\csname urlprefix\endcsname\relax\def\urlprefix{URL }\fi
\expandafter\ifx\csname href\endcsname\relax
  \def\href#1#2{#2} \def\path#1{#1}\fi

\bibitem{MDE}
D.~C. Schmidt, Model-driven engineering, {IEEE} Computer 39~(2) (2006) 25--31.
\newblock \href {https://doi.org/10.1109/MC.2006.58}
  {\path{doi:10.1109/MC.2006.58}}.

\bibitem{MDEDeRun18}
H.~Bruneli{\`{e}}re, R.~Eramo, A.~G{\'{o}}mez, V.~Besnard, J.~Bruel,
  M.~Gogolla, A.~K{\"{a}}stner, A.~Rutle, Model-driven engineering for
  design-runtime interaction in complex systems: Scientific challenges and
  roadmap - report on the mde@derun 2018 workshop, in: {STAF} Workshops, Vol.
  11176 of Lecture Notes in Computer Science, Springer, 2018, pp. 536--543.

\bibitem{CitoLBKMG18}
J.~Cito, P.~Leitner, C.~Bosshard, M.~Knecht, G.~Mazlami, H.~C. Gall,
  {PerformanceHat}: augmenting source code with runtime performance traces in
  the {IDE}, in: Proc. of {ICSE} Companion, 2018, pp. 41--44.

\bibitem{WoodsideFP07}
C.~M. Woodside, G.~Franks, D.~C. Petriu, The future of software performance
  engineering, in: {FOSE} Workshop, {ICSE}, 2007, pp. 171--187.
\newblock \href {https://doi.org/10.1109/FOSE.2007.32}
  {\path{doi:10.1109/FOSE.2007.32}}.

\bibitem{Cortellessa2013}
V.~Cortellessa, Performance antipatterns: State-of-art and future perspectives,
  in: {EPEW} Proc., 2013, pp. 1--6.

\bibitem{Newman15}
S.~Newman, Building Microservices, 1st Edition, O'Reilly Media, Inc., 2015.

\bibitem{ACPET19}
D.~Arcelli, V.~Cortellessa, D.~{Di Pompeo}, R.~Eramo, M.~Tucci, Exploiting
  architecture/runtime model-driven traceability for performance improvement,
  in: {IEEE} International Conference on Software Architecture, {ICSA} 2019,
  Hamburg, Germany, March 25-29, 2019, 2019, pp. 81--90.
\newblock \href {https://doi.org/10.1109/ICSA.2019.00017}
  {\path{doi:10.1109/ICSA.2019.00017}}.

\bibitem{CDEP10}
A.~Cicchetti, D.~Di~Ruscio, R.~Eramo, A.~Pierantonio, {JTL:} {A} bidirectional
  and change propagating transformation language, in: {SLE Proc.}, 2010, pp.
  183--202.
\newblock \href {https://doi.org/10.1007/978-3-642-19440-5\_11}
  {\path{doi:10.1007/978-3-642-19440-5\_11}}.

\bibitem{ARCELLI2018366}
D.~Arcelli, V.~Cortellessa, D.~{Di Pompeo}, {Performance-driven software model
  refactoring}, IST Journal 95 (2018) 366--397.

\bibitem{UML2}
\href{http://www.omg.org/spec/UML/2.5/}{Unified modeling language}, OMG,
  version 2.5 (2015).
\newline\urlprefix\url{http://www.omg.org/spec/UML/2.5/}

\bibitem{MARTE}
\href{http://www.omg.org/omgmarte/}{{A UML profile for MARTE: modeling and
  analysis of real-time embedded systems}}, OMG (2008).
\newline\urlprefix\url{http://www.omg.org/omgmarte/}

\bibitem{Chen18}
L.~Chen, \href{https://doi.org/10.1109/ICSA.2018.00013}{Microservices:
  Architecting for continuous delivery and devops}, in: {IEEE} International
  Conference on Software Architecture, {ICSA} 2018, Seattle, WA, USA, April 30
  - May 4, 2018, 2018, pp. 39--46.
\newblock \href {https://doi.org/10.1109/ICSA.2018.00013}
  {\path{doi:10.1109/ICSA.2018.00013}}.
\newline\urlprefix\url{https://doi.org/10.1109/ICSA.2018.00013}

\bibitem{Cortellessa2011}
V.~Cortellessa, A.~Di~Marco, P.~Inverardi, Model-Based Software Performance
  Analysis, Springer, 2011.
\newblock \href {https://doi.org/10.1007/978-3-642-13621-4}
  {\path{doi:10.1007/978-3-642-13621-4}}.

\bibitem{GL88}
M.~Gelfond, V.~Lifschitz, The stable model semantics for logic programming, in:
  {ICLP}, 1988, pp. 1070--1080.

\bibitem{DLV}
N.~Leone, G.~Pfeifer, W.~Faber, T.~Eiter, G.~Gottlob, S.~Perri, F.~Scarcello,
  The {DLV} system for knowledge representation and reasoning, {TOCL} 7~(3)
  (2006) 499--562.
\newblock \href {https://doi.org/10.1145/1149114.1149117}
  {\path{doi:10.1145/1149114.1149117}}.

\bibitem{EPT18}
R.~Eramo, A.~Pierantonio, M.~Tucci, Improved traceability for bidirectional
  model transformations, in: Proc. of {MDETools} workshop, {MODELS}, Vol. 2245,
  2018, pp. 306--315.

\bibitem{PaigeDKFPOZ11}
R.~F. Paige, N.~Drivalos, D.~S. Kolovos, K.~J. Fernandes, C.~Power, G.~K.
  Olsen, S.~Zschaler, Rigorous identification and encoding of trace-links in
  model-driven engineering, {SOSYM} 10~(4) (2011) 469--487.

\bibitem{Winkler2010}
S.~Winkler, J.~von Pilgrim, A survey of traceability in requirements
  engineering and model-driven development, {SOSYM} 9~(4) (2010) 529--565.
\newblock \href {https://doi.org/10.1007/s10270-009-0145-0}
  {\path{doi:10.1007/s10270-009-0145-0}}.

\bibitem{Kolovos:2010ua}
{Kolovos, Dimitris and Rose, Louis and Paige, Richard and
  Garc{\i}a-Dom{\i}nguez, Antonio}, {The EPSILON book}, Structure, 2010.

\bibitem{Cortellessa:2014cs}
V.~Cortellessa, A.~Di~Marco, C.~Trubiani, {An approach for modeling and
  detecting software performance antipatterns based on first-order logics},
  Software and System Modeling 13~(1) (2014) 391--432.

\bibitem{Smith:2002vl}
C.~U. Smith, L.~G. Williams, {New software performance antipatterns: More ways
  to shoot yourself in the foot}, in: 28th International Computer Measurement
  Group Conference (CMG), 2002, pp. 667--674.

\bibitem{DBLP:journals/sosym/CortellessaMT14}
V.~Cortellessa, A.~D. Marco, C.~Trubiani,
  \href{https://doi.org/10.1007/s10270-012-0246-z}{An approach for modeling and
  detecting software performance antipatterns based on first-order logics},
  Softw. Syst. Model. 13~(1) (2014) 391--432.
\newblock \href {https://doi.org/10.1007/s10270-012-0246-z}
  {\path{doi:10.1007/s10270-012-0246-z}}.
\newline\urlprefix\url{https://doi.org/10.1007/s10270-012-0246-z}

\bibitem{Lazowska:1984ta}
E.~D. Lazowska, J.~Zahorjan, G.~S. Graham, K.~C. Sevcik, {Quantitative system
  performance: computer system analysis using queueing network models},
  Prentice-Hall, Inc., 1984.

\bibitem{DBLP:journals/jacm/ReiserL81}
M.~Reiser, S.~S. Lavenberg,
  \href{https://doi.org/10.1145/322261.322275}{Corrigendum: "mean-value
  analysis of closed multichain queuing networks"}, J. {ACM} 28~(3) (1981) 629.
\newblock \href {https://doi.org/10.1145/322261.322275}
  {\path{doi:10.1145/322261.322275}}.
\newline\urlprefix\url{https://doi.org/10.1145/322261.322275}

\bibitem{DBLP:conf/kbse/ZhouPX0LJD18}
X.~Zhou, X.~Peng, T.~Xie, J.~Sun, W.~Li, C.~Ji, D.~Ding,
  \href{https://doi.org/10.1145/3238147.3240730}{Delta debugging microservice
  systems}, in: M.~Huchard, C.~K{\"{a}}stner, G.~Fraser (Eds.), Proceedings of
  the 33rd {ACM/IEEE} International Conference on Automated Software
  Engineering, {ASE} 2018, Montpellier, France, September 3-7, 2018, {ACM},
  2018, pp. 802--807.
\newblock \href {https://doi.org/10.1145/3238147.3240730}
  {\path{doi:10.1145/3238147.3240730}}.
\newline\urlprefix\url{https://doi.org/10.1145/3238147.3240730}

\bibitem{DBLP:conf/icse/ZhouPX0XJZ18}
X.~Zhou, X.~Peng, T.~Xie, J.~Sun, C.~Xu, C.~Ji, W.~Zhao,
  \href{https://doi.org/10.1145/3183440.3194991}{Benchmarking microservice
  systems for software engineering research}, in: M.~Chaudron, I.~Crnkovic,
  M.~Chechik, M.~Harman (Eds.), Proceedings of the 40th International
  Conference on Software Engineering: Companion Proceeedings, {ICSE} 2018,
  Gothenburg, Sweden, May 27 - June 03, 2018, {ACM}, 2018, pp. 323--324.
\newblock \href {https://doi.org/10.1145/3183440.3194991}
  {\path{doi:10.1145/3183440.3194991}}.
\newline\urlprefix\url{https://doi.org/10.1145/3183440.3194991}

\bibitem{DBLP:conf/sigsoft/Zhou0X0JLXH19}
X.~Zhou, X.~Peng, T.~Xie, J.~Sun, C.~Ji, D.~Liu, Q.~Xiang, C.~He,
  \href{https://doi.org/10.1145/3338906.3338961}{Latent error prediction and
  fault localization for microservice applications by learning from system
  trace logs}, in: M.~Dumas, D.~Pfahl, S.~Apel, A.~Russo (Eds.), Proceedings of
  the {ACM} Joint Meeting on European Software Engineering Conference and
  Symposium on the Foundations of Software Engineering, {ESEC/SIGSOFT} {FSE}
  2019, Tallinn, Estonia, August 26-30, 2019, {ACM}, 2019, pp. 683--694.
\newblock \href {https://doi.org/10.1145/3338906.3338961}
  {\path{doi:10.1145/3338906.3338961}}.
\newline\urlprefix\url{https://doi.org/10.1145/3338906.3338961}

\bibitem{DBLP:conf/staf/Pompeo0CE19}
D.~{Di Pompeo}, M.~Tucci, A.~Celi, R.~Eramo,
  \href{http://ceur-ws.org/Vol-2405/06\_paper.pdf}{A microservice reference
  case study for design-runtime interaction in {MDE}}, in: A.~Bagnato,
  H.~Bruneli{\`{e}}re, L.~Burgue{\~{n}}o, R.~Eramo, A.~G{\'{o}}mez (Eds.),
  {STAF} 2019 Co-Located Events Joint Proceedings: 1st Junior Researcher
  Community Event, 2nd International Workshop on Model-Driven Engineering for
  Design-Runtime Interaction in Complex Systems, and 1st Research Project
  Showcase Workshop co-located with Software Technologies: Applications and
  Foundations {(STAF} 2019), Eindhoven, The Netherlands, July 15 - 19, 2019,
  Vol. 2405 of {CEUR} Workshop Proceedings, CEUR-WS.org, 2019, pp. 23--32.
\newline\urlprefix\url{http://ceur-ws.org/Vol-2405/06\_paper.pdf}

\bibitem{DBLP:conf/fase/ArcelliCT15}
D.~Arcelli, V.~Cortellessa, C.~Trubiani,
  \href{https://doi.org/10.1007/978-3-662-46675-9\_10}{Performance-based
  software model refactoring in fuzzy contexts}, in: A.~Egyed, I.~Schaefer
  (Eds.), Fundamental Approaches to Software Engineering - 18th International
  Conference, {FASE} 2015, Held as Part of the European Joint Conferences on
  Theory and Practice of Software, {ETAPS} 2015, London, UK, April 11-18, 2015.
  Proceedings, Vol. 9033 of Lecture Notes in Computer Science, Springer, 2015,
  pp. 149--164.
\newblock \href {https://doi.org/10.1007/978-3-662-46675-9\_10}
  {\path{doi:10.1007/978-3-662-46675-9\_10}}.
\newline\urlprefix\url{https://doi.org/10.1007/978-3-662-46675-9\_10}

\bibitem{Kounev21}
J.~Grohmann, M.~Straesser, A.~Chalbani, S.~Eismann, Y.~Arian, N.~Herbst,
  N.~Peretz, S.~Kounev, Suanming: Explainable prediction of performance
  degradations in microservice applications, in: {ICPE} '21: {ACM/SPEC}
  International Conference on Performance Engineering, Virtual Event, France,
  April 19-21, 2021, {ACM}, 2021, pp. 165--176.

\bibitem{DBLP:journals/infsof/TrubianiBHAK18}
C.~Trubiani, A.~Bran, A.~van Hoorn, A.~Avritzer, H.~Knoche, Exploiting load
  testing and profiling for performance antipattern detection, {IST} Journal 95
  (2018) 329--345.

\bibitem{DBLP:conf/models/PorterMG16}
J.~Porter, D.~A. Menasc{\'{e}}, H.~Gomaa, Desarm: {A} decentralized mechanism
  for discovering software architecture models at runtime in distributed
  systems, in: {Proc. of Models@run.time workshop, MODELS}, Vol. 1742, 2016,
  pp. 43--51.

\bibitem{DBLP:conf/models/GomaaA17}
H.~Gomaa, E.~Albassam, Run-time software architectural models for adaptation,
  recovery and evolution, in: {Proc. of Models@run.time workshop, MODELS}, Vol.
  2019, 2017, pp. 193--200.

\bibitem{Altamimi:2016:PAR:3049877.3049899}
T.~Altamimi, M.~H. Zargari, D.~C. Petriu, Performance analysis roundtrip:
  Automatic generation of performance models and results feedback using
  cross-model trace links, in: Proc. of {CASCON}, 2016, pp. 208--217.

\bibitem{Vogele2018}
C.~V{\"o}gele, A.~van Hoorn, E.~Schulz, W.~Hasselbring, H.~Krcmar, {{WESSBAS:}
  extraction of probabilistic workload specifications for load testing and
  performance prediction - a model-driven approach for session-based
  application systems}, {SOSYM} 17~(2) (2018) 443--477.

\bibitem{Mazkatli20}
M.~{Mazkatli}, D.~{Monschein}, J.~{Grohmann}, A.~{Koziolek}, Incremental
  calibration of architectural performance models with parametric dependencies,
  in: 2020 IEEE International Conference on Software Architecture (ICSA), 2020,
  pp. 23--34.
\newblock \href {https://doi.org/10.1109/ICSA47634.2020.00011}
  {\path{doi:10.1109/ICSA47634.2020.00011}}.

\bibitem{Heinrich20}
R.~Heinrich, \href{https://doi.org/10.1016/j.jss.2020.110722}{Architectural
  runtime models for integrating runtime observations and component-based
  models}, J. Syst. Softw. 169 (2020) 110722.
\newblock \href {https://doi.org/10.1016/j.jss.2020.110722}
  {\path{doi:10.1016/j.jss.2020.110722}}.
\newline\urlprefix\url{https://doi.org/10.1016/j.jss.2020.110722}

\bibitem{RademacherSZ17}
F.~Rademacher, S.~Sachweh, A.~Z{\"{u}}ndorf, Differences between model-driven
  development of service-oriented and microservice architecture, in: 2017
  {IEEE} International Conference on Software Architecture Workshops, {ICSA}
  Workshops 2017, 2017, pp. 38--45.
\newblock \href {https://doi.org/10.1109/ICSAW.2017.32}
  {\path{doi:10.1109/ICSAW.2017.32}}.

\bibitem{Derakhshanmanesh16}
M.~Derakhshanmanesh, M.~Grieger, Model-integrating microservices: {A} vision
  paper, in: Gemeinsamer Tagungsband der Workshops der Tagung Software
  Engineering 2016 {(SE} 2016), 2016, pp. 142--147.

\bibitem{DBLP:books/sp/19/Weyns19}
D.~Weyns, \href{https://doi.org/10.1007/978-3-030-00262-6\_11}{Software
  engineering of self-adaptive systems}, in: S.~Cha, R.~N. Taylor, K.~C. Kang
  (Eds.), Handbook of Software Engineering, Springer, 2019, pp. 399--443.
\newblock \href {https://doi.org/10.1007/978-3-030-00262-6\_11}
  {\path{doi:10.1007/978-3-030-00262-6\_11}}.
\newline\urlprefix\url{https://doi.org/10.1007/978-3-030-00262-6\_11}

\bibitem{modelsruntime}
G.~{Blair}, N.~{Bencomo}, R.~B. {France}, Models@run.time, Computer 42~(10)
  (2009) 22--27.

\bibitem{Bencomo2019}
N.~Bencomo, S.~G{\"o}tz, H.~Song, Models@run.time: a guided tour of the state
  of the art and research challenges, Software {\&} Systems Modeling (2019).

\bibitem{DullmannH17}
T.~F. D{\"{u}}llmann, A.~van Hoorn, Model-driven generation of microservice
  architectures for benchmarking performance and resilience engineering
  approaches, in: Companion Proceedings of the 8th {ACM/SPEC} on International
  Conference on Performance Engineering, {ICPE} 2017, 2017, pp. 171--172.
\newblock \href {https://doi.org/10.1145/3053600.3053627}
  {\path{doi:10.1145/3053600.3053627}}.

\bibitem{Prieto2017}
M.~Z{\'u}{\~{n}}iga-Prieto, E.~Insfran, S.~Abrah{\~a}o, C.~Cano-Genoves,
  Automation of the incremental integration of microservices architectures, in:
  Complexity in Information Systems Development, 2017, pp. 51--68.

\bibitem{Sampaio2019}
A.~R. Sampaio, J.~Rubin, I.~Beschastnikh, N.~S. Rosa, Improving
  microservice-based applications with runtime placement adaptation, Journal of
  Internet Services and Applications 10~(1) (2019) 4.
\newblock \href {https://doi.org/10.1186/s13174-019-0104-0}
  {\path{doi:10.1186/s13174-019-0104-0}}.

\end{thebibliography}

\end{document}